\documentclass[12pt]{article}
\pdfoutput=1
\usepackage{jheppub}
\usepackage[skip=2pt,font=footnotesize]{caption}
\usepackage[utf8]{inputenc}
\usepackage{verbatim}
\usepackage{amsmath}
\usepackage{amssymb}
\usepackage{amsthm}
\usepackage{amsfonts}
\usepackage{makecell}
\usepackage{tabularx}
\usepackage{comment}
\usepackage{subfigure}
\usepackage{float}
\usepackage{graphicx}

\usepackage{hyperref}
\usepackage{psfrag}
\usepackage{comment}

\def\1{{\rm 1-loop}}

\newcommand{\Oc}{\mathcal {O}}

\newcommand{\F}{{}_2F_1}
\newcommand\numberthis{\addtocounter{equation}{1}\tag{\theequation}}
\usepackage{braket}

\usepackage[T1]{fontenc}
\inputencoding{utf8}

\usepackage{enumitem}

\newcommand*\pFq[6][8]{%
  \begingroup % only local assignments
  \pFqmuskip=#1mu\relax
  \mathcode`\,=\string"8000
  \begingroup\lccode`\~=`\,
  \lowercase{\endgroup\let~}\pFqcomma
  {}_{#2}F_{#3}{\left[\genfrac..{0pt}{}{#4}{#5};#6\right]}%
  \endgroup
}
\newcommand{\pFqcomma}{\mskip\pFqmuskip}
\makeatletter
\renewcommand{\@maketitle}{
\newpage
 \begin{center}%
  {\large\bfseries \@title \par}%
 \end{center}%
 \par} \makeatother
\numberwithin{equation}{section}

\begin{document}

\begin{flushright}
    \footnotesize
    CALT-TH 2024-024
\end{flushright}

\title{Correlators of Worldline Proper Length}

\abstract{A classical observer can measure elapsed proper time along their worldline. When observers are coupled to a system with internal correlations, measurements of elapsed time may inherit these correlations. We show that derivatives of the on-shell action with respect to worldline mass compute correlation functions of worldline proper length at tree level. We study worldlines coupled to a scalar field. We calculate the length-length two-point function and find it arises from correlated path fluctuations. As an application, we propose that the logarithm of local correlators serves as a generating function of length correlators, which generalizes the on-shell action prescription. Using this proposal, we extract AdS worldline observables from local CFT correlators as computed by Witten diagrams. We briefly discuss extensions to gravity, interferometers, and the holographic encoding of observer time.}

\author{ Allic Sivaramakrishnan
\let\thefootnote\relax\footnote{\texttt{allic@caltech.edu}}}

\affiliation{Walter Burke Institute for Theoretical Physics,\\ California Institute of Technology, Pasadena, CA 91125}

%\affiliation{Caltech}{Walter Burke Institute for Theoretical Physics,
%    California Institute of Technology, Pasadena, CA 91125 }

\maketitle

\section{Introduction}
\label{section1}

An idealized classical observer can measure elapsed proper time by consulting their clock twice. We can model this observer as a massive point particle that travels between two spacetime points, which we specify as input. The observer's elapsed time is then the proper length of the particle's worldline.

It is less clear how to describe these elapsed time measurements beyond the deterministic classical regime. Nevertheless, a quantity we can measure once can also have non-trivial correlation in a suitable setting. Suppose multiple observers are coupled to a common system that displays internal correlations. This system may be classical and noisy or a quantum degree of freedom, for example a quantum field in a spatially entangled state. In this setup, measurements made by the observers can inherit correlations from the common system. In particular, when the observers measure their respective elapsed times, they may discover these times are correlated. One can calculate correlators of worldline observables\footnote{By worldline observable, we mean a formal quantity built only from information accessible along the worldline. This should not be confused with what an actual experiment measures in practice.} built from local operators integrated over worldlines using standard QFT methods, at least in certain regimes \cite{Witten23a,Witten23b}. However, less is known about correlation functions of elapsed proper time.

In this paper, we study correlation functions of worldline proper length. We work mostly in Euclidean signature for simplicity, and so the preceding Lorentzian setup serves only as motivation. We primarily study worldlines coupled to a scalar field.

In gravitational systems, certain measurements can depend crucially on the nature of the observer.\footnote{One example of recent interest is entropy. See \cite{GiacominiCB17,AliAhmadGHLS21,ChandrasekaranLPW22,HoehnKM23,JensenSS23,DeVuystEHK24} and references therein.} Because of this, studying general worldline observables in detail may help us construct gravitational observables. Observer dependence in gravity has a long history, and this topic is sometimes known as the study of quantum reference frames when the observers are dynamical and possess quantum degrees of freedom (see \cite{BartlettRS06, VanrietveldeHGC18} for reviews and \cite{ChandrasekaranLPW22,DeVuystEHK24} for recent connections to algebras of observables). An observer-centric approach must be able to account for interactions between observers and their environment or with the systems they measure. This means it is necessary to describe time measurements made along non-inertial worldlines carrying stress-energy rather than along abstractly-defined geodesics. Defining time via measurements at some energy scale may help us refine effective field theory methods that elucidate how quantum gravity effects modify notions of locality and geometry. For example, loop corrections may probe the ultraviolet properties of time, but also of distances, horizons, and causality.

Many topics ranging from theory to experiment involve studying worldlines and worldline observables. We briefly review certain results that may connect to the ideas in this paper.

Worldline observables appear in various experimental and experimentally-oriented work on gravitational physics. The phase shift measured by matter-wave interferometers can be expressed as the on-shell action of a point particle travelling in a closed loop  \cite{StoreyC94,DimopoulosGHK08}. The electromagnetic Aharonov-Bohm effect is a classic example of this, and the gravitational Aharonov-Bohm effect was also recently measured \cite{OverstreetACKK2022}. Other examples of quantum matter interacting via gravity, including in atom interferometers, also involve worldline configurations \cite{AsenbaumOKBHK16, OverstreetCKAKG22,BengyatDAC23, CarneyCGMPST22,CarneyMT21}. Recent proposals for quantum gravity signatures in laser interferometers, \cite{VZ1,VZ2,VZ3,Pixellon,BanksZ21} and \cite{ParikhWZ20a, ParikhWZ20b,ParikhWZ20c}, amount to predictions for how length fluctuations induced by quantum gravity effects are correlated. Recent work on observational signatures of black hole mergers that LIGO may detect uses worldline-centric approaches (for example, see \cite{ GoldbergerR04,Porto16, KalinP20,MogullPS20,CheungPRSW23,CheungPRSW24} and references therein).

Worldline time has long been studied in a fully quantum-mechanical setting, for example in Unruh-DeWitt detectors. Defining time as an operator requires care, but this topic appears in recent work on crossed product algebra constructions, relational observables, and quantum reference frames, for example see \cite{HoehnSL19,HoehnSL20, Giacomini21,ChandrasekaranLPW22,Witten23a,Witten23b,JensenSS23}. Experimental determination of time-of-flight during tunnelling processes may capture worldline time in a highly quantum regime \cite{Davies04,RamosDRS19,SuzukiU22}. In quantum mechanics, time operators were originally studied in \cite{SusskindG64} for applications to quantum clocks, a length operator for crystals was discussed in \cite{Schrodinger54}, a quantum ruler was constructed in \cite{WangGNB23}, and the connection between measurement resolution and the length of a quantum-mechanical path was studied in \cite{AbbottW79}.

If observer dependence is truly an important feature of quantum gravity, we may want to understand how observers are encoded holographically in a dual non-gravitational theory. This topic has been addressed recently using the Anti-de Sitter/Conformal Field Theory correspondence (AdS/CFT), the best-studied example of holography. A simple first step is to identify the holographic dual of worldline proper length. The canonical procedure is to work in a limit in which the geodesic approximation applies to propagation in AdS, in which case CFT correlators encode the length of geodesics or geodesic networks in AdS. Using this approach, the time an observer takes to reach the black hole singularity can be extracted from boundary correlators (see \cite{GrinbergM20, HorowitzLQZ23, CeplakLPV24} for recent progress).\footnote{This approach should be contrasted with that of \cite{LeutheusserL21,LeutheusserL22}, which focuses on algebras of observables.} The work of \cite{JafferisL20,deBoerJL22} studied measurements made along non-geodesic worldlines and the connection between boundary modular time and bulk worldline time. Timelike entanglement entropy on the boundary encodes bulk time along piecewise-extremal surfaces via an analytically-continued version of the Ryu-Takayanagi prescription \cite{DoiHMTT22,DoiHMTT23}. Length operators have also been studied in holographic settings. A bulk length operator for null geodesics and its boundary dual were obtained in the Regge limit of boundary correlators \cite{Afkhami-JeddiHKT17}. Bulk length operators were recently studied in low dimensions, specifically in two-dimensional gravity \cite{IliesiuLLMM24} and in a geometric dual of double-scaled SYK \cite{AlmheiriGH24}. While progress on extracting bulk worldline data from the boundary has accelerated recently, this subject remains inchoate.

Many of the worldline observables we reviewed are local quantities integrated along the entire worldline, of which worldline proper time is a simple example. However, correlators of these nonlocal operators are in general less well studied than local correlators. Moreover, the background these operators live on, the worldline, is dynamical and backreaction changes the worldline's geometry. Nevertheless, worldline observables are such ubiquitous objects that we may want a convenient generating function for their multipoint correlation functions.

The main goal of this paper is to present a method for calculating correlation functions of worldline proper length. The method is simple: derivatives of the on-shell action $\bar{S}$ with respect to worldline mass $m$ define classical nonlinear response functions of worldline proper length. Computed perturbatively, these are interpreted as connected tree-level contributions to correlation functions of proper length. $\bar{S}$ therefore serves as a generating function of these correlators. This method is ultimately equivalent to the standard approach to computing correlation functions (or nonlinear response functions) with nonzero sources. Here, we use the observation that $m$ functions as a source for proper length in the worldline action. Similar statements apply to derivatives with respect to coupling constants $\lambda$.

In more detail, consider $n$ point particles of mass $m_i$ coupled to a scalar field $\psi$ via $\lambda_i f_i(\psi)$. Suppose the $i$-th worldline connects spacetime points $x_{i,a}$ and $x_{i,b}$. The action is
\begin{equation}
S(m_1,\cdots,m_n) = S_{\psi}+\sum_{i=1}^n \int d\tau_i \left(m_i\sqrt{\dot{x}_i^\mu(\tau_i)\dot{x}_i^\nu(\tau_i)g_{\mu \nu}(x_i(\tau_i))}+ \lambda_i f_i(\psi(x_i(\tau_i)))\right),
\end{equation}
where $S_{\psi}$ is the action for $\psi$. We show that the on-shell proper length $\bar{L}_i$ of worldline $i$ can be extracted from the on-shell action $\bar{S}$ as
\begin{equation}
\frac{d}{dm_i} \bar{S}(m_1,\cdots,m_n) = \int d\tau_i \sqrt{\dot{\bar{x}}_i^\mu(\tau_i)\dot{\bar{x}}_i^\nu(\tau_i)g_{\mu \nu}(\bar{x}_i(\tau_i))} 
= \bar{L}_i.
\label{Intro1ptClaim}
\end{equation}
Classical response functions of proper length are then by definition
\begin{equation}
\braket{L_1 L_2}_c = -\frac{d}{dm_2} \bar{L}_1 = -\frac{d}{dm_1} \frac{d}{dm_2}\bar{S}(m_1,\cdots,m_n),
\label{Intro2ptDefinition}
\end{equation}
and similarly for $n$-point functions, 
\begin{equation}
\braket{L_1 \cdots L_n}_c = -\left(\prod_{i=1}^n -\frac{d}{dm_i} \right) \bar{S}(m_1,\cdots,m_n).
\label{IntronptDefinition}
\end{equation}
The quantity \eqref{IntronptDefinition} is the main object of study in this paper. We have used braket notation for convenience and the subscript $c$ stands for connected.

To understand the intuition behind \eqref{IntronptDefinition}, recall that in QFT, consistency with the classical limit requires that functional derivatives $\delta /\delta J(x)$ of the on-shell action $\bar{S}[J]$ with $S[J]= S[0] + \int d^{D}x J(x) \Oc(x)$ define connected correlation functions of $\Oc$ even when $J(x)$ remains nonzero. In classical physics, such quantities are known as nonlinear response functions. We emphasize that here $J(x)$ is not taken small, and so these functions are the responses to perturbations of the form $J(x) \rightarrow J(x) + \delta J(x)$. 

To study properties of proper length correlators, we compute $\braket{L_1 L_2}_c$ to lowest order in $\lambda_i$ as a simple but nontrivial example. We show that $\braket{L_1 L_2}_c$ arises from correlated path fluctuations to all orders in $\lambda_i$, which provides a conceptual consistency check.

Because tree diagrams and certain perturbative solutions to the equations of motion are equivalent,\footnote{This well-known fact is the essence of Berends-Giele recursion, in which tree-level scattering amplitudes are extracted from perturbative solutions to the equations of motion.} the correlators $\braket{L_1 \cdots L_n}_c$ defined by \eqref{IntronptDefinition} are a prediction for tree-level correlators of the proper time measurements made along worldlines coupled to a quantum field $\psi$. We briefly discuss potential extensions to gravity and a toy model of matter-wave interferometer measurements.

Although our primary goal is to present \eqref{IntronptDefinition}, we explore the following application. We propose that $\eqref{IntronptDefinition}$ can be rephrased in terms of correlation functions of local operators in the semiclassical regime, which we denote by $\hbar \rightarrow 0$. For example,
\begin{align*}
\hbar \rightarrow 0: ~~~~\braket{L_1 L_2}_c &= \frac{d}{dm_1} \frac{d}{dm_2} \log Z(m_1,m_2),
\\
 Z(m_1,m_2) &\equiv \braket{\phi_1(x_{1,a})\phi_1(x_{1,b}) \phi_2(x_{2,a}) \phi_2(x_{2,b})},
 \numberthis
\label{Intro2ptCorrelator}
\end{align*}
where $\phi_i$ is a scalar field of mass $m_i$. In the semiclassical regime, correlators of $\phi_i$ encode $\bar{S}$ perturbatively in $\lambda$ via saddle-point approximation to the well-known worldline representation of two-point functions (\cite{BernKosower,Strassler92,StrasslerThesis}). In this way, \eqref{Intro2ptCorrelator} embeds the first-quantized description of length correlators in a second-quantized formulation.

One may ask whether there is a fully quantum formulation of length correlators, in other words, a rigorous definition of all matrix elements of length beyond the semiclassical regime. We briefly discuss how \eqref{Intro2ptCorrelator} with $\hbar \neq 0$ appears to furnish one such definition, for example
\begin{equation}
\hbar \neq 0:  ~~~~~\braket{L_1 L_2}_c = \frac{d}{dm_1} \frac{d}{dm_2} \log Z(m_1,m_2).
\end{equation}
In this sense, correlators of local operators may serve as generating functions for correlators of proper length in the fully quantum regime, modulo certain potential subtleties we mention. Similarly, $d/d\lambda_i$ derivatives would define correlators of worldline operators $\int d\tau_i f_i(\psi(x_i(\tau_i)))$.

Finally, we apply a version of \eqref{Intro2ptCorrelator} to CFT correlators and compute what we argue are bulk worldline observables. In this context, we propose that for large scaling dimensions $\Delta$ of the external operators, bulk proper length is extracted via $d/d\Delta$ derivatives of the logarithm of CFT correlators.

We now summarize the organization of this paper. The content is divided into two sections.

Section \ref{section2} contains the length correlator prescription in terms of the on-shell action. In Section \ref{L}, we give the prescription for extracting $\bar{L}$ from the on-shell action. Then in Section \ref{proof}, we justify this prescription and explain how similar statements apply for derivatives with respect to any explicit parameter in the action. In Section \ref{LL}, we compute $\braket{L_1 L_2}_c$ as an example. In Section \ref{sec2applications}, we briefly discuss applications to gravity and interferometers. 

In Section \ref{section3}, we propose extensions of the on-shell action prescription to correlation functions and AdS/CFT. In Section \ref{CorrelatorLift}, we lift the on-shell action method to correlators of local operators and also discuss application to the fully quantum regime. In Section \ref{CorrelatorExamples}, we give a number of simple examples. Finally, in Section \ref{AdSCFT}, we apply a version of this local correlator proposal to AdS/CFT, compute worldline observables from Witten diagrams, and discuss possible CFT interpretations. 

We conclude with future directions in Section \ref{futures}, where we discuss possible applications of the methods presented here.

\section{Length Correlators From the On-Shell Action}
\label{section2}

In this section, we study the prescription for defining and computing correlators of proper length. For simplicity, we mostly study proper length in flat Euclidean spacetime with worldlines coupled to a scalar field. Generalizations to proper time in Lorentzian spacetimes, interactions with other fields, and curved backgrounds appear straightforward. 

\subsection{Worldline Proper Length $\braket{L}$}
\label{L}
We give a prescription for extracting worldline proper length from a generating function.  Consider a relativistic point particle of mass $m$ at position $x$ that is coupled to a scalar field $\psi$ via the interaction term $\lambda f(\psi(x))$, where $f$ is some analytic function. We impose boundary conditions $x(\tau_1)=x_1, x(\tau_2)=x_2$ on $x$ with worldline time $\tau \in [\tau_1,\tau_2]$. The parameter $\tau$ is unphysical, as are $\tau_1,\tau_2$. Quantities $\tau_1,\tau_2$ and $x_1,x_2$ are independent of $m, \lambda$, and specified as inputs. The action for this theory is
\begin{equation}
S = \int_{\tau_1}^{\tau_2} d\tau \left(m\sqrt{\eta_{\mu\nu}\dot{x}^\mu(\tau)\dot{x}^\nu(\tau) }+ \lambda f(\psi(x(\tau)))\right) + S_{\psi}.
\label{WorldlineActionSingle}
\end{equation}
The first two terms are the worldline action and $S_{\psi} = \int d^D x \mathcal{L}_\psi$ is the action for $\psi$ living in $D$ spacetime dimensions. The field $\psi$ may be dynamical and free, $\mathcal{L}_\psi = \frac{1}{2}(\partial \psi)^2$, dynamical and fully interacting, for example $\mathcal{L}_\psi = \frac{1}{2}(\partial \psi)^2 + \frac{\lambda}{3}\psi^3$, or non-dynamical, $\mathcal{L}_\psi = 0$. 

We study $S$ evaluated on-shell, which we denote by $\bar{S} \equiv S[\bar{x},\bar{\psi}]$, the action evaluated on the support of the equations of motion for $x,\psi$. When featured in a path integral for paths $x(\tau)$ and the field $\psi$, $\bar{S}$ is the saddle that dominates in the semiclassical regime, assuming a single saddle dominates. In a path integral, the einbein form of the worldline action is used. However, because we restrict our attention to the on-shell setup here, we use the square-root action. 

We emphasize a crucial point, that we have assumed a Lagrangian and a well-behaved variational principle. We do not, however, assume a path-integral formulation or quantum theory otherwise defined from which the on-shell action descends via a classical limit. In fact, we will discuss later what may be required to extend our results to a quantum theory.

The on-shell value of $x$, $\bar{x}$, solves the equation of motion
\begin{equation}
\partial_\tau^2 \bar{x}^\mu = \frac{\lambda}{m} f'(\bar{\psi}(\bar{x})) \frac{\partial \bar{\psi}(\bar{x})}{\partial \bar{x}_\mu},
\end{equation}
where the on-shell value of $\psi$, $\bar{\psi}$, solves the equation of motion for $\psi$. We have suppressed the $\tau$ argument for brevity above. As we will show in Section \ref{proof},
\begin{equation}
\frac{d}{dm}\bar{S}= \int_{\tau_1}^{\tau_2} d\tau \sqrt{\eta_{\mu\nu}\dot{\bar{x}}^\mu(\tau)\dot{\bar{x}}^\nu(\tau) } = \bar{L} \equiv \braket{L},
\label{OnePointFunctionProposal}
\end{equation}
where $\bar{L}$ is the proper length of the on-shell worldline. Note that the on-shell action is the sum of both the worldline and $\psi$ actions evaluated on shell. The length computed this way is classical as we began from the action of a classical theory.

We chose a single scalar $\psi$ with a particular coupling for definiteness, but \eqref{OnePointFunctionProposal} is much more general. If $m$ appears explicitly only in front of the proper length term, then irrespective of what other terms appear in the action or whether the theory can be consistently quantized, \eqref{OnePointFunctionProposal} applies.

\subsection{Correlators From Derivatives of the On-Shell Action}
\label{proof}

In this section, we derive \eqref{OnePointFunctionProposal} and explain how the same idea defines higher-point correlators of length. We will aim to be pedagogical and also set some notation.

First, recall the standard approach to computing connected correlators of operators $\mathcal{O}$ in a quantum theory via the generating functional $Z[J]$,
\begin{align*}
\braket{\Oc(x_1)} &= -\frac{\delta}{\delta J(x_1)} \log Z[J],
\\
\braket{\Oc(x_1)\Oc(x_2)}_c &= \frac{\delta}{\delta J(x_2)}\frac{\delta}{\delta J(x_1)} \log Z[J] = -\frac{\delta}{\delta J(x_2)} \braket{\Oc(x_1)},
\numberthis
\label{QFTCorrelatorPrescription}
\end{align*}
with
\begin{equation}
Z[J] = \int \mathcal{D}[\phi] e^{-\frac{1}{\hbar}S[\phi,J]},
~~~~~~~
S[\phi,J]=S[\phi,0]+\int d^D x J(x)\Oc(x),
\end{equation}
where $\phi$ stands for the fundamental degrees of freedom in the path integral and $\Oc$ is a function of $\phi$, for example $\Oc = \phi^n$. Here, $J(x)$ acts as a source for $\Oc(x)$. We use quantum field theory for definiteness, although analogous statements apply in quantum mechanics.

After computing \eqref{QFTCorrelatorPrescription}, we commonly set $J(x)=0$ to obtain the correlators in the $J(x) = 0$ theory. However, another standard choice is to retain the nonzero source. The procedure \eqref{QFTCorrelatorPrescription} then defines correlators in the $J(x) \neq 0 $ theory. We will consider this $J(x) \neq 0$ case throughout.

In the semiclassical regime, the path integral localizes onto its saddle and therefore $\phi$ localizes onto $\bar{\phi}$, a solution to the equations of motion. This is also known as going ``on shell,'' which we indicate using barred notation.\footnote{See \cite{MeltzerS20, OnShellCorrelators} for clarification of how this notion of on shell compares to what appears in unitarity cuts.} 

On shell, \eqref{QFTCorrelatorPrescription} is evaluated with $\log Z[J] = -\bar{S}[J]/\hbar$ with $\bar{S}[J] \equiv S[\bar{\phi},J]$. We denote the semiclassical regime, which is $\bar{S}[J]/\hbar \gg 1$, as $\hbar \rightarrow 0$ for simplicity. In the semiclassical regime, $\braket{\Oc}$ is extracted from the on-shell action $\bar{S}$ as
\begin{equation}
\hbar \rightarrow 0: ~~~~~\frac{\delta}{\delta J(x_1)} \bar{S}[J]  = \braket{\Oc(x_1)} = \bar{\Oc}(x_1),
\label{Sec2OnePointFunction}
\end{equation}
where we have assumed that, to leading order in the $\hbar \rightarrow 0 $ limit, the one-point function is equal to the on-shell value, $\braket{\Oc(x_1)}=\bar{\Oc}(x_1)$. It should be emphasized that these two may differ by an overall power of $\hbar$ that we will omit here, or equivalently remove by including the appropriate prefactor with $\frac{\delta}{\delta J(x)}$. As we will explain, this assumption is manifestly true in perturbation theory at tree level, which is our focus here. With this understanding, we will use braket notation $\braket{\Oc}$ in place of $\bar{\Oc}$ in the semiclassical regime for notational convenience. 

The two-point function is
\begin{equation}
\hbar \rightarrow 0: 
~~~~~~
\braket{\Oc(x_1)\Oc(x_2)}_c =-\frac{\delta}{\delta J(x_2)}\braket{\Oc(x_1)}
=
-\frac{\delta}{\delta J(x_1)}\frac{\delta}{\delta J(x_2)} \bar{S}[J],
\label{Sec2TwoPointFunction}
\end{equation}
and similarly for higher points. In the classical theory there is no ensemble to sum over, and so as we will explain, these quantities are not interpreted as connected correlators in the classical theory. Everything we have discussed so far is simply the standard method of recovering classical physics from the semiclassical regime of the path integral, which is sometimes also called the classical limit for this reason. 

If we applied the relation \eqref{Sec2OnePointFunction} to the worldline action \eqref{WorldlineActionSingle} with $J(x) \rightarrow m$, we would find that proper length $\int ds$ plays the role of $\Oc$ and so our claim \eqref{OnePointFunctionProposal} would then follow. However, strictly speaking, $\eqref{Sec2OnePointFunction}$ relied on the existence of a path integral and an assumption about recovering classical physics via saddle point approximation. Because the starting point of our main claim \eqref{OnePointFunctionProposal} is the on-shell action and not a quantum generating function, neither of these arguments directly prove \eqref{OnePointFunctionProposal}. Nevertheless, consistency with the classical limit of quantum field theory does strongly suggest that $\frac{\delta}{\delta J(x_1)} \bar{S}[J]  = \bar{\Oc}(x_1)$ may be a general property of on-shell actions, and it would be surprising if a proof did not exist. Such a proof would automatically imply our main claim \eqref{OnePointFunctionProposal} regardless of whether $\bar{S}$ descends from a quantum generating function.

Fortunately, direct proof of \eqref{Sec2OnePointFunction} is straightforward. We will prove a more general statement. For definiteness, consider a field $\phi(x)$ with Lagrangian density $\mathcal{L}(\lambda,\phi,\partial \phi)$. The parameter $\lambda$ stands for all the explicit parameters in the Lagrangian, including masses, coupling constants, and explicit functions like $J(x)$. We also assume that the Dirichlet boundary conditions imposed on the spacetime manifold's boundary $\Sigma$ are independent of $\lambda$, or $\frac{d}{d\lambda}\bar{\phi}(x_B)=0$ where $x_B \in \Sigma$. We assume the location of $\Sigma$ is also independent of $\lambda$. In this setup, $\partial_\mu \bar{\phi}(x_B)$ will generically depend on $\lambda$. With $\bar{\phi}=\bar{\phi}(x,\lambda)$,
\begin{align*}
\frac{d}{d\lambda} \bar{S}(\lambda)
&=
\frac{d}{d\lambda} \int d^D x \mathcal{L}(\lambda,\bar{\phi},\partial \bar{\phi}) 
\\
&=
\int d^Dx \left( 
\frac{\partial}{\partial \lambda} 
+ \frac{d\bar{\phi}}{d\lambda} \frac{\delta}{\delta \phi} 
+\frac{d\partial_\mu \bar{\phi}}{d\lambda} \frac{\delta}{\delta \partial_\mu \phi}
\right)\mathcal{L}(\lambda,\phi,\partial \phi)\bigg|_{\phi=\bar{\phi}(x,\lambda)}.
\numberthis
\end{align*}
Integrating the third term by parts produces a boundary term proportional to $\frac{d\bar{\phi}(x_B)}{d\lambda}=0$, and so
\begin{equation}
\frac{d}{d\lambda} \bar{S}(\lambda)
=
\int d^Dx 
\left( 
\frac{\partial}{\partial \lambda} \mathcal{L}(\lambda,\phi,\partial \phi)
+\frac{d\bar{\phi}}{d\lambda} \left( 
\frac{\delta}{\delta  \phi}
-
\partial_\mu \frac{\delta}{\delta \partial_\mu \phi}
\right)
\mathcal{L}(\lambda,\phi,\partial \phi)
\right)
\bigg|_{\phi=\bar{\phi}(x,\lambda)}.
\end{equation}
As $\bar{\phi}$ satisfies the Euler-Lagrange equations,
\begin{equation}
\frac{d}{d\lambda} \bar{S}(\lambda)
=
\int d^Dx 
\frac{\partial}{\partial \lambda} \mathcal{L}(\lambda,\phi,\partial \phi)\bigg|_{\phi=\bar{\phi}(x,\lambda)}
,
\end{equation}
which proves \eqref{Sec2OnePointFunction} and therefore also \eqref{OnePointFunctionProposal}. In other words, $d/d\lambda$ acting on $\bar{S}$ extracts the term expected from the $d/d\lambda$ derivative of $S$ off shell, but this term is evaluated on shell.

The proof we gave applies to any theory with a Lagrangian and well-defined variational principle. However, not all quantities obtained by taking derivatives with respect to parameters have simple physical interpretations. Proper length, \eqref{OnePointFunctionProposal}, is a notable exception.

If the boundary conditions depend on $\lambda$, one may be able to take a similar approach, either by subtracting off the additional contribution or by analytically continuing the boundary conditions to be independent of $\lambda$, acting with $\frac{d}{d\lambda}$, and then imposing the desired boundary conditions.

For completeness, we give an elementary example of \eqref{Sec2OnePointFunction}. Consider the simple harmonic oscillator, $\mathcal{L} = \frac{1}{2}m \dot{x}^2 - \frac{1}{2}k x^2$. With boundary conditions $x(\tau_1)=x_1, x(\tau_2)=x_2 $, the solution is
\begin{equation}
\bar{x}(\tau)=\frac{x_2-x_1 \cos(\omega (\tau_2-\tau_1))}{\sin (\omega (\tau_2-\tau_1))}\sin(\omega (\tau-\tau_1)) + x_1 \cos(\omega (\tau-\tau_1)).
\end{equation}
One can check that indeed
\begin{align*}
\int_{\tau_1}^{\tau_2} d\tau \frac{1}{2}m \dot{\bar{x}}^2(\tau) &= m\frac{d}{dm} \bar{S}(m,k),
\\
\int_{\tau_1}^{\tau_2} d\tau \frac{1}{2} k \bar{x}^2(\tau) &= -k \frac{d}{dk} \bar{S}(m,k).
\numberthis
\end{align*}
One may also compute higher-point functions in this way when there are multiple masses connected by springs, or in the many exactly solvable systems in classical mechanics.

The higher-point quantities extracted via derivatives of parameters, for example \eqref{Sec2TwoPointFunction}, are a type of classical nonlinear response function. Concretely, consider the action $S[\phi,J_i] = S[\phi,0]+\sum_{j=1}^n \int d^Dx \Oc_j(x)J_j(x)$. Under perturbations $J_i \rightarrow J_i + \delta J_i$,
\begin{align*}
\bar{S}[J_i] \rightarrow  \bar{S}[J_i] &+\left(\sum_{j} \delta J_j(x_j) \frac{\delta}{\delta J_j(x_j)}\right)\bar{S}[J_i] 
\\
&+ 
\left(\sum_{j\neq k } \delta J_j(x_j) \delta J_k(x_k) \frac{\delta}{\delta J_j(x_j)}\frac{\delta}{\delta J_k(x_k)}\right)
\bar{S}[J_i]+
\cdots 
\\
&
+ \left(\prod_j \delta J_j(x_j) \frac{\delta}{\delta J_j(x_j)} \right)\bar{S}[J_i]+\cdots.
\numberthis
\end{align*}
The coefficients of the perturbations $\delta J_j$ at quadratic order and higher are known as response functions,\footnote{In this sense, $\bar{\Oc}$ is a linear response function of the on-shell action.} as they describe changes in one-point functions under linear perturbations $\delta J_j$ or nonlinear perturbations $\delta J_j \delta J_k$,
\begin{align*}
\bar{S}[J_i] \rightarrow  \bar{S}[J_i] &+\sum_j \delta J_j(x_j) \braket{\Oc(x_j)}[J_i] 
+ 
\sum_{j\neq k} \delta J_j(x_j)\delta J_k(x_k) \braket{\Oc_j(x_j)\Oc_k(x_k)}_c[J_i] 
+
\cdots 
\\
&
+ \left(\prod_j \delta J_j(x_j)\right) \braket{\Oc_1(x_1) \cdots \Oc_n(x_n)}_c[J_i] +\cdots.
\numberthis
\end{align*}
One can think of the source $J(x)$ as being controlled by a knob that an experimentalist can dial. The value of an external field is one common example of $J(x)$. In optics and magnetism, response functions are known as susceptibilities. 

Response functions are typically studied as perturbations around the $J_i=0$ theory. Computations may be easier here than in the $J_i \neq 0 $ case, and in practice we often seek to quantify how the system changes when the sources $J_i$ are turned on. When $J_i \neq 0$, the response functions capture how sensitive classical quantities are to changes in the $J_i$ already present.

Classical response functions computed perturbatively at weak coupling can also be interpreted as tree-level correlation functions in the corresponding weakly-coupled quantum theory. It is a well-known fact that tree diagrams can be extracted from the perturbative solutions to the classical equations of motion at weak coupling with the appropriate zeroth-order seeds. In the context of scattering amplitudes, this is known as Berends-Giele recursion, but a similar statement holds for correlation functions (see \cite{OnShellCorrelators} for detailed discussion). 

We can therefore compute tree-level correlators in a quantum theory without ever constructing the quantum operators. We simply take derivatives of $\bar{S}$, which is fully determined by classical physics. We will refer to the resulting nonlinear response functions computed perturbatively as tree-level correlation functions for simplicity. Even though there is no ensemble in the classical case, these response functions and correlation functions differ only by overall powers of $\hbar$, as explained earlier. 

In short, the equivalence between classical response functions and tree-level correlators will allow us to calculate the tree-level length correlators of a quantum theory from the on-shell action. We can entirely avoid constructing a time operator in worldline quantum mechanics or rigorously defining a path-integral prescription, both of which may involve subtleties.

\subsection{Correlators of Proper Length $\braket{LL}$}
\label{LL}

As discussed in Section \ref{proof}, the approach taken in Section \ref{L} also defines higher-point correlation functions of proper length at tree level. The computations in this section illustrate this prescription and will provide a conceptual consistency check of interpreting these correlators as tree-level processes in a quantum theory.

We now state the prescription for tree-level correlation functions of worldline proper length. Consider $n$ distinct point particles with masses $m_i$. Suppose the $i$-th particle is described by worldline $i$ connecting spacetime points $x_{2i-1},x_{2i}$.
\begin{equation}
\braket{L_1(x_1,x_2)\cdots L_n(x_{2n-1},x_{2n})}_c = -\prod_{i=1}^n \left(-\frac{d}{dm_i}\right) \bar{S}(x_1,x_2,\cdots,x_{2n-1},x_{2n}).
\label{Sec2LLGeneralPrescription}
\end{equation}
where $\bar{S}(x_1,x_2,\cdots,x_{2n-1},x_{2n})$ is the on-shell action for the $n$ point particles and any other systems they are coupled to. Note that acting with powers of a single $d/dm_i$ computes correlators along the same worldline, although we do not study this case.

In this section, we primarily study an example of \eqref{Sec2LLGeneralPrescription},
\begin{equation}
\braket{L_1 L_2}_c = -\frac{d^2 \bar{S}}{dm_1 dm_2 }.
\end{equation}
Consider two worldlines $x_1, x_2$ coupled to a scalar field $\psi(x)$ that has no self interactions,
\begin{equation}
S= \sum_{i=1}^2 \int d\tau_i \left(m_i\sqrt{\eta_{\mu\nu}\dot{x}_i^\mu(\tau_i)\dot{x}_i^\nu(\tau_i) }+ \lambda_i f_i(\psi(x_i(\tau_i)))\right) 
+
\frac{1}{2}\int d^Dx (\partial \psi(x))^2.
\label{Sec2TwoWorldlineAction}
\end{equation}
We will mostly suppress the argument $\tau_i$ to simplify notation. 

Our goal is to compute $\braket{L_1 L_2}_c$ to lowest order in $\lambda_i$ by solving the equations of motion perturbatively in $\lambda_i$. Taking $f_i(\psi)=\psi$ for simplicity, the equations of motion are
\begin{align*}
&\partial_{\tau_i}^2 x^\mu_i =\frac{\lambda_i}{m_i} \frac{\partial\psi(x_i)}{\partial x_{i,\mu}},
\\
&\square \psi(x) = \sum_{i=1}^2 \lambda_i \int d\tau_i \delta^{(D)}(x,x_i).
\numberthis
\end{align*}
We omit the bars on the on-shell quantities $\bar{x}_i, \bar{\psi}$ in this section to reduce notational clutter. $\delta^{(D)}(x,x')$ denotes the Dirac delta function. Expanding $x_i, \psi$ about $\lambda_i =0$,
\begin{align*}
x_i^\mu &= \sum_{p,q=0} x_i^{(p,q),\mu} \lambda_1^p \lambda_2^q,
\\
\psi(x) &= \psi^{(0)}(x) + \Oc(\lambda_i).
\numberthis
\end{align*}
The exact solutions to the equations of motion can then be written as
\begin{align*}
x^\mu_i(\tau_i)&=x_i^{(0,0),\mu}(\tau_i)+\frac{\lambda_i}{m_i} \int d\tau_i' G_x(\tau_i,\tau_i')\frac{\partial}{\partial x_{i',\mu}}
\left(
\psi^{(0)}(x_{i'})
+
\sum_{j=1}^2 \lambda_j \int d\tau_j G_\psi(x_{i'},x_j)
\right)
,
\\
\psi(x) &= \psi^{(0)}(x) +\sum_{i=1}^2 \lambda_i \int d\tau_i G_\psi(x,x_i),
\numberthis
\end{align*}
where $x_{i'} \equiv x_i(\tau_i')$. The above implicitly defines $\bar{x}_i,\bar{\psi}$ in perturbation theory. The Green's functions and homogeneous solutions obey
\begin{alignat*}{2}
\square_{x} G_\psi(x,x') &= \delta^{(D)}(x,x'),~~~~~~~~~~~~ &\quad  \square_x \psi^{(0)}(x) &= 0, 
\\
\partial_{\tau}^2 G_x(\tau,\tau') &= \delta(\tau,\tau'),~~~~~~~~~~~~ &\quad  \partial_{\tau_i}^2 x_i^{(0,0),\mu}&=0.
\numberthis
\end{alignat*}
The worldline propagator $G_x(\tau,\tau')$ represents a correction to the worldline path, or in other words, signifies a path fluctuation. $\psi^{(0)}$ is the value of $\psi$ in the absence of the worldlines, and a nonzero background value for $\psi$ around which we expand. $\psi^{(0)}$ and $x^{(0,0)}$ serve as seeds for tree diagrams, which are generated by perturbatively solving the equations of motion. In the language of Feynman diagrams, $\psi^{(0)}$ and $x^{(0,0)}$ are on-shell external legs (see \cite{OnShellCorrelators} for further discussion). 

To determine which orders in perturbation theory are needed to compute $\braket{L_1 L_2}_c$ to lowest order, it will be useful to rewrite $\bar{S}$ by using the $\psi$ equation of motion,
\begin{equation}
\bar{S}=
\sum_{i=1}^2 \int d\tau_i \left( m_i\sqrt{\eta_{\mu\nu}\dot{x}_i^\mu\dot{x}_i^\nu }
+ 
\frac{1}{2}\lambda_i\psi^{(0)}(x_i) \right)
+
\frac{1}{2}\sum_{i,j=1}^2 \lambda_i \lambda_j \int d\tau_i d\tau_j G_\psi(x_i,x_j).
\end{equation}
It may be more computationally efficient to evaluate the entire on-shell action at once and then act with the relevant derivatives. However, we will instead compute the terms of the on-shell action individually in order to better interpret features of length correlators. We write
\begin{equation}
\braket{L_1 L_2}_c = -\frac{d^2 \bar{S}}{dm_1 dm_2}
=-(T_1+T_2+T_3),
\end{equation}
with
\begin{align*}
T_1=&
\frac{d^2}{dm_1 dm_2}
\sum_{i=1}^2 m_i\int d\tau_i \sqrt{\eta_{\mu\nu}\dot{x}_i^\mu\dot{x}_i^\nu },
\\
T_2=&
\frac{1}{2}\frac{d^2}{dm_1 dm_2}
\sum_{i=1}^2 \int d\tau_i  \lambda_i\psi^{(0)}(x_i),
\\
T_3=&
\frac{1}{2}\frac{d^2}{dm_1 dm_2}
\sum_{i,j=1}^2 \lambda_i \lambda_j \int d\tau_i d\tau_j G_\psi(x_i,x_j).
\numberthis
\label{3terms}
\end{align*}
Next we will identify the perturbative solutions to the equations of motion that contribute to the terms above. This can be done systematically by tracking the factors of $m_i$ via the following logic. No term in the off-shell action $S$ contains both $m_1, m_2$. Therefore, the terms in $\bar{S}$ that survive $\frac{d^2}{dm_1 dm_2}$ will need the additional factors of $m_i$ that come from imposing the equations of motion for $x_i$. For example, to survive the derivative, a term in $S$ that contains only $m_1,x_1$ needs to couple to a path fluctuation of $x_2$, because as we see from the equations of motion, this is the only way to recruit a factor of $m_2$. This coupling therefore turns the term that was proportional to $m_1$ off shell into one proportional to $m_1^p m_2^q$ with $p,q \neq 0$ on shell. This term survives $\frac{d^2}{dm_1 dm_2}$ and so it contributes to $\braket{L_1 L_2}_c$. The equations of motion also reveal that powers of $1/m_i$ count the number of propagators $G_x(\tau_i,\tau_{i}')$, in other words are a measure of how many path fluctuations a process involves. Although we consider arbitrary $m_i$ here, powers of $1/m_i$ in the large-$m_i$ expansion do indeed track path fluctuations in this way. 

We now compute the necessary perturbative solutions to the equations of motion, in other words the tree-level worldline Feynman diagrams, that we will need. This amounts to performing Berends-Giele recursion with multiple coupling constants. We will proceed order by order in $\lambda_i$.

We will consider $x_1^{(p,q)}$ without loss of generality and discuss the interpretation of each relevant term in the expansion. The lowest-order correction we need is 
\begin{equation}
\lambda_1 x_1^{(1,0),\mu}(\tau_1)=
\frac{\lambda_1}{m_1} \int d\tau_1' G_x(\tau_1,\tau_1')\frac{\partial}{\partial x_{1',\mu}^{(0,0)}}
\psi^{(0)}(x_{1'}^{(0,0)}).
\end{equation}
In $x_1^{(1,0),\mu}$ we see that the background value of $\psi^{(0)}$ exerts a force on the worldline and alters its trajectory. The equation of motion for $x_1$ implies that this path perturbation comes with one power of $1/m_1$ and one worldline propagator $G_x$.

The other first-order correction is $x_1^{(0,1),\mu} =0$, as $x_1$ does not couple to $x_2$ at this order.

Next we consider the $\Oc(\lambda_i^2, \lambda_i \lambda_j)$ terms. We obtain these terms by perturbatively expanding the $x_1$ equation of motion. First,
\begin{align*}
\lambda_1^2 x^{(2,0),\mu}_1(\tau_1) =& \frac{\lambda_1^2}{m_1}\int d\tau_1' G_x(\tau_1,\tau_1') \int d\tau_1''  \frac{\partial}{\partial x_{1',\mu}^{(0,0)}}  G_\psi(x^{(0,0)}_{1'},x_{1''}^{(0,0)})
\\&+
\frac{\lambda_1^2}{m_1} \int d\tau_1' G_x(\tau_1,\tau_1') \left( x_{1'}^{(1,0)} \cdot \frac{\partial}{\partial x_{1'}^{(0,0)}}\right) \frac{\partial}{\partial x_{1',\mu}^{(0,0)}}
\psi^{(0)}(x_{1'}^{(0,0)}).
\numberthis
\end{align*}
The first term is a path correction due $x_1$ emitting and then absorbing a $\psi$, in other words self interaction via $\psi$. Because $\psi$ is sourced by the worldline and not the background $\psi^{(0)}$, this term has an additional power of $\lambda_1$ compared to $x_1^{(1,0)}$. The second term arises because $\psi^{(0)}$ kicks the trajectory and so $\psi^{(0)}$ now couples to the new location.

It is clear that $x_1^{(0,2)} = 0$, and in fact that $x_1^{(0,n)} = 0$ for all $n \geq 1$. This is because $\psi$ is not self-interacting and so $x_1$ fluctuates only due to its interaction with $\psi$, which is controlled by $\lambda_1$.

Next, 
\begin{equation}
\lambda_1 \lambda_2  x^{(1,1),\mu}_1(\tau_1) = \frac{\lambda_1\lambda_2}{m_1}\int d\tau_1' G_x(\tau_1,\tau_1') \int d\tau_2  \frac{\partial}{\partial x_{1',\mu}^{(0,0)}}  G_\psi(x^{(0,0)}_{1'},x_2^{(0,0)})
\end{equation}
describes a coupling between the two worldlines. This process is the change in $x_1$ due to $\psi$-mediated interaction with the unperturbed worldline $x_2$. This is the first fully-connected diagram we encounter.

We will also need
\begin{equation}
\lambda_1 \lambda_2^2 x_1^{(1,2),\mu}(\tau_1) = \frac{\lambda_1\lambda_2^2}{m_1} \int d\tau_1' G_x(\tau_1,\tau_{1'})
  \int d\tau_2  \left( x_2^{(0,1)} \cdot \frac{\partial}{\partial x_{2}^{(0,0),\nu}}\right) \frac{\partial}{ \partial x_{1',\mu}^{(0,0)} } G_\psi(x_{1'}^{(0,0)},x_2^{(0,0)}).
\end{equation}
This describes $x_2$ exerting a force on $x_1$, but where $x_2$ has itself changed in response to the field $\psi^{(0)}$ as captured by $x_2^{(0,1)}$. This diagram therefore can be interpreted as a fluctuation in $x_2$ coupling to a fluctuation in $x_1$. As advertised, the first appearance of a $1/m_2$ factor arises from the fluctuation of $x_2$, which is $x_2^{(0,1)}$.

By contrast, $x_1^{(2,1)}$ does not contain a $1/m_2$. We will not need $x_1^{(2,1)}$ or any terms that are higher order in $\lambda_i$ to compute $\braket{L_1 L_2}_c$ to lowest order. 

We can now compute the lowest-order contribution to each term in $\bar{S}$. Term $T_1$ in \eqref{3terms} is
\begin{align*}
T_1 =& \frac{d^2}{dm_1 dm_2}
\sum_{i=1}^n \int d\tau_i m_i\sqrt{\eta_{\mu\nu}\dot{x}_i^\mu\dot{x}_i^\nu }
\\
&
\approx
\lambda_1^2 \lambda_2^2 \frac{d^2}{dm_1 dm_2} \left(
m_1 \int d\tau_1 ~\dot{x}_1^{(1,0)}\cdot \dot{x}_1^{(1,2)}
+
m_2 \int d\tau_2 ~\dot{x}_2^{(0,1)}\cdot \dot{x}_2^{(2,1)}
\right).
\numberthis
\end{align*}
We showed previously that $\dot{x}_1^{(1,0)}\cdot \dot{x}_1^{(1,2)} \propto 1/(m_1^2 m_2)$. For notational simplicity, we have parametrized $x_i$ such that $(\dot{x}_i^{(0,0)})^2=1$. Because $x_1^{(0,0)}$ extremizes the free worldline action, only second-order variations in the path will contribute. This is why even though $\dot{x}_1^{(2,2)} \cdot \dot{x}_1^{(0,0)}$ may contain the same powers of $\lambda_i, m_i$ as the term above, $\dot{x}_1^{(2,2)} \cdot \dot{x}_1^{(0,0)}$ is a term in the perturbative expansion of a first-order path variation and therefore does not contribute.

As $T_2$ does not contain an explicit $m_1$ or $m_2$, its lowest-order term arises from the lowest-order correction to $x_i$ that is proportional to $1/(m_1 m_2)$. For example, the contribution from the $\psi^{(0)}(x_1)$ term in $T_2$ is
\begin{align*}
&\frac{\lambda_1}{2} \frac{d^2}{dm_1 dm_2}  \int d\tau_1 \psi^{(0)}(x_1) 
\approx
 \frac{\lambda_1^2 \lambda_2^2}{4} \frac{d^2}{dm_1 dm_2}\int d\tau_1 
\frac{d}{d\lambda_1}\frac{d^2}{d\lambda_2^2}
\psi^{(0)}(x_1) \bigg|_{\lambda_i=0}
\\
&~~=\frac{\lambda_1^2 \lambda_2^2}{4}  \frac{d^2}{dm_1 dm_2} \int d\tau_1 
\frac{d}{d\lambda_1}
\left(
 \frac{\partial ^2 x_1}{\partial \lambda_2^2}\cdot \frac{d}{d x_{1}} 
+
\left(
\frac{\partial x_1^\mu}{\partial \lambda_2}  
\frac{\partial x_1^\nu}{\partial \lambda_2} 
\frac{d}{dx_1^\mu}
\frac{d}{dx_1^\nu} 
\right)
\right)\psi^{(0)}(x_1)
\bigg|_{\lambda_i=0}.
\numberthis
\end{align*}
From the perturbative expansion of $x_1$, we see that 
\begin{equation}
\frac{\partial^n x_1^\mu}{\partial \lambda_2^n} \bigg|_{\lambda_i = 0} = 0 ~~~~~~~~\text{for} ~~~~~~~~ n \geq 1,
\end{equation}
which means that the only surviving term arises from $\frac{\partial^3 x_1}{\partial \lambda_1 \partial \lambda_2^2}\big|_{\lambda_i =0 } = x_1^{(1,2)}$. This term is proportional to $1/(m_1 m_2)$ and so it contributes to $T_2$.
\begin{align*}
T_2=&\frac{d^2}{dm_1 dm_2} \sum_i \frac{\lambda_i}{2} \int d\tau_i \psi^{(0)}(x_i) 
\\
\approx&
\frac{\lambda_1^2 \lambda_2^2}{4} \frac{d^2}{dm_1 dm_2} 
\Bigg(
\int d\tau_1 
 \left( x^{(1,2)}_1 \cdot\frac{\partial}{\partial x_{1}^{(0,0)}}\right) \psi^{(0)}(x_1^{(0,0)})
 \\
 &~~~~~~~~~~~~~~~~~~~~~~
+
\int d\tau_2 
 \left( x^{(2,1)}_2 \cdot\frac{\partial}{\partial x_{2}^{(0,0)}}\right) \psi^{(0)}(x_2^{(0,0)}) 
 \Bigg)
 .
\numberthis
\end{align*}
Finally, the lowest-order contribution to $T_3$ is 
\begin{align*}
T_3 &= \frac{\lambda_1\lambda_2 }{2} \frac{d^2}{dm_1 dm_2} \int d\tau_1 d\tau_2 G_\psi(x_1,x_2)
\\
&\approx
 \frac{\lambda_1^2\lambda_2^2}{2} \frac{d^2}{dm_1 dm_2}\int d\tau_1 d\tau_2 
 x_1^{(1,0),\mu} 
 x_2^{(0,1),\nu}  
 \frac{\partial^2}{\partial x_1^{(0,0),\mu} \partial x_2^{(0,0),\nu}} G_\psi(x_1^{(0,0)},x_2^{(0,0)}).
\numberthis
\end{align*}
We dropped the $i = j$ terms in $\sum_{i,j=1}^2 \lambda_i \lambda_j \int d\tau_i d\tau_j G_\psi(x_i,x_j)$ because these contribute terms higher order in $\lambda_i$ than the $i\neq j$ terms. This is because, for example, to obtain $1/m_2$ from $G_\psi(x_1,x_{1'})$, one needs up to $x_{1}^{(1,2)}$, which is $\Oc(\lambda_1 \lambda_2^2)$.

This concludes the calculation of $\braket{L_1 L_2}_c$. Specifically, we have computed the position-space integrand. We do not evaluate the integrals here. Computing position-space diagrams in closed form is somewhat non-trivial, beyond the scope of this work, and is more easily done in specific applications or kinematic limits. Computations may be simpler in momentum space than position space.

\begin{figure}[H]
\centering
\includegraphics[width=\textwidth]{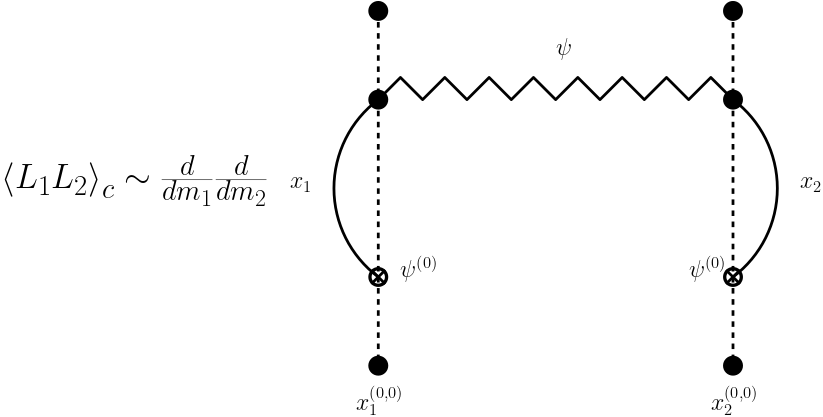}
\caption{
This and similar tree-level processes generate all nonzero contributions to $\braket{L_1 L_2}_c$ at tree level to all orders in $\lambda_i$, and at higher multiplicity. This diagram represents the correlation of a path fluctuation of $x_1$ with a path fluctuation of $x_2$ due to correlations inherited from $\psi$. The solid lines are worldline propagators $G_x(\tau,\tau')$, the zig-zag line is the $\psi$ propagator $G_\psi(x_1^{(0,0)},x_2^{(0,0)})$, and the dotted lines are the unperturbed solutions $x_1^{(0,0)}, x_2^{(0,0)}$ along which worldline vertices are integrated. The coupling of $x_i$ to the background value of the field $\psi^{(0)}$, depicted by $\otimes$, contributes a factor of $\lambda_i/m_i$. The vertex joining $\psi$ and $x_i$ contributes $\lambda_i$. The diagram is then $\Oc(\lambda_1^2 \lambda_2^2)$.}
\label{fig1}
\end{figure}

While the $T_i$ have different forms, they share one important feature: they all contain a diagram with a worldline propagator $G_x(\tau_i,\tau_i')$ for each worldline and are fully connected via a $G_\psi(x_1,x_2)$ propagator between worldlines. See Figure \ref{fig1} for a diagrammatic representation of $T_3$, in which the fluctuations of $x_1,x_2$ are both sourced by $\psi^{(0)}$. Terms $T_1,T_2$ contain $x_1^{(1,2)}$ and $x_2^{(2,1)}$, which are similar to $T_3$ but where instead only one of the worldline fluctuations is directly sourced by $\psi^{(0)}$. 

The diagrams we encountered represent correlated path fluctuations, and so are precisely the diagrams expected to contribute to $\braket{L_1 L_2}_c$. One can also view these diagrams as the response of a path fluctuation of one worldline to the path fluctuation in the other worldline. As higher-order terms are determined recursively by lower order ones, the diagrams we find here determine the higher-order terms in $\braket{L_1 L_2}_c$. We also expect similar statements apply to higher-point correlators of $L_i$. In summary, we find that the equations of motion imply that every tree-level contribution to $\braket{L_1 \cdots L_n}_c$ contains a diagram of correlated path fluctuations, which is a conceptual consistency check of our proposal.

Finally, note that we can also compute the two-point function of the integrated worldline operator $\psi(\gamma_i) \equiv \int d\tau_i \psi(x_i)$ where $x_i$ lives on the worldline $\gamma_i$. Setting $\psi^{(0)} =0$ and taking $m_i \rightarrow \infty $ for simplicity, to lowest order in $\lambda_i$, we have
\begin{equation}
\frac{d}{d\lambda_1} \frac{d}{d\lambda_2}\bar{S}  = 
\int d\tau_1 d\tau_2 G_\psi(x_1^{(0,0)},x_2^{(0,0)})=\braket{\psi(\gamma_1) \psi(\gamma_2)},
\end{equation}
as expected. At finite values of $m_i$, there are additional diagrams that correspond to fluctuations in $x_i$. The two-point function of $\psi$ inserted at fixed spacetime points is independent of the dynamics of $\gamma_i$, but $\psi(\gamma_i)$ and its correlators receive contributions from effects that deform $\gamma_i$.

\subsection{Applications}
\label{sec2applications}

The setup of worldlines coupled to a scalar field illustrates certain basic features of proper length correlators. We comment briefly on possible applications to more physically relevant settings: gravity and interferometry.

\subsubsection{Gravity}

In this section, we discuss rudimentary features of $\braket{L_1 L_2}_c$ for worldlines coupled to linearized gravity. The main purpose of this section is to connect the prescription in Section \ref{LL} to the canonical quantity studied as a length operator, which is an integral of $h^{\mu\nu}$ over a worldline $\gamma$, or schematically $\int_\gamma h$. We show that to leading order in $G_N$, $\braket{L_1 L_2}_c$  takes the form $\int_{\gamma_1 }\int_{\gamma_2}G_{h}(x_1,x_2)$, as expected from the connected two-point function of $\int_\gamma h$ with graviton propagator $G_h$. 

As before, we fix the worldline endpoints at coordinates chosen in the reference frame of, for example, an inertial observer. We emphasize however that addressing the diffeomorphism-invariance of correlators defined in this way is beyond the scope of this paper. 

We expand $g_{\mu \nu} = \eta_{\mu \nu}+ \kappa h_{\mu\nu}$ with $\kappa^2 = 32 \pi G_N$. The action is
\begin{equation}
S =  \sum_{i=1}^2 m_i\int d\tau_i \sqrt{\dot{x}_1^\mu \dot{x}_1^\nu (\eta_{\mu\nu}+\kappa h_{\mu\nu}(x_1))} +\frac{2}{\kappa^2} \int d^D x  R,
\end{equation}
where we leave the linearization of $R = R(\eta_{\mu\nu}+\kappa h_{\mu\nu})$ implicit. To lowest order in $\kappa$,
\begin{equation}
S
=
\sum_{i=1}^2 m_i\int d\tau_i \sqrt{\dot{x}_i^\mu \dot{x}_i^\nu \eta_{\mu\nu}}
+
\frac{1}{2}\kappa \sum_{i=1}^2 m_i \int d\tau_i  \dot{x}_i^\mu \dot{x}_i^\nu h_{\mu\nu}(x_i)  
+ 
\frac{2}{\kappa^2} \int d^D x  R.
\label{GravAction}
\end{equation}
The equations of motion are
\begin{align*}
&\frac{d^2 x^\mu_i}{d\tau^2_i} + \frac{\kappa}{2}\eta^{\mu \lambda}\left(\partial_\rho h_{\lambda\sigma}(x_i) + \partial_\sigma h_{\lambda\rho}(x_i) -\partial_\lambda h_{\rho \sigma}(x_i)\right)\frac{dx^\rho_i}{d\tau} \frac{dx^\sigma_i}{d\tau}=0,
\\
&h^{\mu\nu}(x) = \kappa \int d^Dx' G_h^{\mu\nu,\rho\sigma}(x,x')  T_{\rho\sigma}(x'),
\\
&
T^{\mu\nu}(x)=\frac{1}{2}\sum_{i=1}^2 m_i \int d\tau_i \dot{x}^\mu_i \dot{x}^\nu_i \delta^{(D)}(x-x_i),
\numberthis
\end{align*}
where $G_h^{\mu\nu,\rho\sigma}(x,x')$ is the graviton propagator. We have set the solution to the homogenous equation of motion for $h$ to zero for simplicity, $h^{(0),\mu\nu} = 0$.

Next, we compute the leading $\kappa$ correction to $\braket{L_1 L_2}_c$. We expand $x_i$ as
\begin{equation}
x_i^\mu = x_i^{(0),\mu}+ \kappa^2 x_i^{(2),\mu},
\end{equation}
where $x_i^{(0)}$ solves the geodesic equation with $g_{\mu\nu} = \eta_{\mu\nu}$, and $x_i^{(1)}=0$. 

We now study the terms in the linearized off-shell action \eqref{GravAction}. The leading correction to the first term, $\int d\tau_i \sqrt{\dot{x}_i^{\mu}\dot{x}_i^{\nu}\eta_{\mu\nu}}$, arises at $\Oc(\kappa^4)$. Because $x^{(0)}$ extremizes this quantity, the lowest-order correction arises from lowest-order corrections to each $\dot{x}_i$ in the square root, and each is $\Oc(\kappa^2)$. A similar argument appeared in the scalar example studied earlier.

The lowest-order correction to the second term in \eqref{GravAction} arises from a correction to $h_{\mu\nu}(x_i)$. From the equations of motion, we see this correction is $\Oc(\kappa^2)$. We therefore have
\begin{equation}
\braket{L_1 L_2}_c = -\frac{d}{dm_1} \frac{d}{dm_2} \bar{S} =
\kappa^2 \int d\tau_1  d\tau_2 \dot{x}_{1,\mu}^{(0)}  \dot{x}_{1,\nu}^{(0)} G_h^{\mu \nu,\rho\sigma}(x_1,x_2) \dot{x}_{2,\rho}^{(0)}  \dot{x}_{2,\sigma}^{(0)},
\end{equation}
up to an overall numerical factor we have not tracked. This matches the connected two-point function of what is commonly referred to as the length operator,
\begin{align*}
\braket{L_1 L_2}_c &\approx \braket{\hat{L}_1 \hat{L}_2}_c,
\\
\hat{L}_i \approx
\int d\tau_i \sqrt{\dot{x}_i^{(0),\mu} \dot{x}_i^{(0),\nu}\eta_{\mu \nu}} &+ \frac{1}{2} \kappa \int d\tau_i \dot{x}_i^{(0),\mu} \dot{x}_i^{(0),\nu} \hat{h}_{\mu \nu}(x_i).
\numberthis
\end{align*}
Note that $\hat{L}_i$ contains a $c$-number term, or a term proportional to the identity, that does not have an obvious operator interpretation.\footnote{A similar statement applies to the vacuum modular Hamiltonian for subregions in conformal field theory. We discuss a possible connection in the context of holographic correlators in Section \ref{CorrelatorEntropy}.}

\subsubsection{Correlators of Interferometer Phase Shifts}

Our approach to length correlators may provide predictions for experiments. Here, we give a toy example that illustrates how our method computes correlation functions of phase shifts captured by matter-wave interferometer setups when the probe is coupled to a quantum field. Similar statements may apply to atom interferometers.

The observable in a matter-wave interferometer is the relative phase between massive probes traversing different paths. Our discussion follows \cite{StoreyC94, DimopoulosGHK08}. Under time evolution, the probe accrues phase according to a semiclassical approximation to the propagator. The phase $\varphi_i$ accumulated over a path $\mathcal{C}_i$ is the on-shell action $\bar{S}$ of the effectively pointlike probe, $\varphi_i = \bar{S}[\mathcal{C}_i]$. The phase shift $\Phi_{12}=\varphi_1-\varphi_2$ can be written as the on-shell action for a closed loop in spacetime, $\mathcal{C}_{12}$, which is the curve $\mathcal{C}_1$ followed by the path-reversed curve $\mathcal{C}_2$,
\begin{equation}
\Phi_{12} = \bar{S}[\mathcal{C}_{12}]=\int_{\mathcal{C}_{12}} d\tau \mathcal{L}(\bar{x}(\tau)).
\end{equation}
In this way, the phase shift can also be thought of as a generalized version of holonomy.

A well-known example is the holonomy of a charged particle in a background gauge field, otherwise known as the Aharonov-Bohm phase. The holonomy is captured by the expectation value of a Wilson loop,
\begin{equation}
\mathcal{W}[\mathcal{C}_{12}] = \mathcal{P}\left( e^{i q \int_{\mathcal{C}_{12}}  A^\mu dx_\mu}\right), ~~~~~ \braket{\mathcal{W}[\mathcal{C}_{12}]} = e^{i \Phi_{12}},
\end{equation}
where $q$ is the charge of the probe particle, $A^\mu$ is a background gauge field, and the Aharonov-Bohm phase is $\Phi_{12}$. To compare the on-shell action and $\Phi_{12}$, we begin with the action of a point particle coupled to a background gauge field,
\begin{equation}
S = \int d\tau \left( m \sqrt{\dot{x}^2(\tau)} + q A^\mu(x(\tau)) \dot{x}_\mu(\tau) \right).
\end{equation}
Choosing curves $\mathcal{C}_{1}, \mathcal{C}_2$ to have identical geometry, the contribution to $\bar{S}[\mathcal{C}_{12}]$ from path differences is $m \int_{\mathcal{C}_{12}} d\tau \sqrt{\dot{x}^2(\tau)} = 0$, and so
\begin{equation}
\bar{S}[\mathcal{C}_{12}] = q \int_{\mathcal{C}_{12}}d\tau A^\mu(x(\tau)) \dot{x}_\mu(\tau) = q\int_{\mathcal{C}_{12}}A^\mu(x)dx_\mu =\Phi_{12},
\end{equation}
as expected. Wilson lines can also be obtained from the gauge theory action by including a term $J^\mu A_\mu$, where $J$ has delta-function support along $\mathcal{C}_{i}$. In the Aharonov-Bohm setup most commonly considered, path fluctuations are not present, but they can easily be included.

Using the ideas in Section \eqref{proof}, we can calculate an additional quantity that may be in principle observable: $n$-point correlation functions of the interferometer phase shift when the probe is coupled to a scalar field $\psi$, which can be thought of as a toy model of $A^{\mu}$. Because $\psi$ has spacetime correlations, we expect interferometer measurements performed in different spacetime regions may be correlated. 

As an illustrative example, we consider the theory of two worldlines in \eqref{Sec2TwoWorldlineAction} with coupling $f_i(\psi) = \psi$. Taking the large-$m_i$ limit implies the paths receive no corrections and so we can engineer $\mathcal{C}_i$ to be curves of identical shape. The phase shift therefore arises only from interaction with $\psi$.

Suppose the curves $\mathcal{C}_{12}, \mathcal{C}_{12'}$ have the same geometry but are related by a spacetime translation. This represents two identical phase shift measurements occuring at different locations in spacetime. Working to lowest order in $\lambda_i$ and in the large-$m_i$ limit, the on-shell action for the closed curves is
\begin{equation}
\bar{S}[\mathcal{C}_{12},\mathcal{C}_{12'}] = \lambda_1\lambda_2 \int_{\mathcal{C}_{12},\mathcal{C}_{12'}} ds_{12} ds_{12'}  G_\psi(x(s_{12}),x(s_{12'})).
\end{equation}
The integrated form of the above quantity is not particularly illuminating for our purposes. However, suppose $\mathcal{C}_{12}$ has total length $\ell$ and $\mathcal{C}_{12}, \mathcal{C}_{12'}$ are separated by $L$. In the $L \gg l$ limit, the connected two-point function of phase shifts scales in $L$ as
\begin{equation}
\braket{\Phi_{12}\Phi_{12'}}_c = \lambda_1 \lambda_2 \frac{d}{d\lambda_1}\frac{d}{d\lambda_2} \bar{S}[\mathcal{C}_{12},\mathcal{C}_{12'}] \approx \lambda_1 \lambda_2  \frac{l^2}{L^{(D-2)/2}}e^{-m_\psi L},
\end{equation}
up to overall polynomial factors of $m_\psi$, the mass of $\psi$, and overall numerical factors. The massless case can be found by setting $m_\psi = 0$ above. The decay of this correlation in $L$ distinguishes massless and massive $\psi$ for large enough $L$.

We expect a similar result applies to gauge theories. In principle, the two-point function of the Aharonov-Bohm phase may probe the QED vacuum. Whether similar statements hold in gravity is unclear, but a natural object to consider may be the so-called scalar gravitational Wilson line (for example, see \cite{AlawadhiBWW21}),
\begin{equation}
\mathcal{W}_{\mathcal{C}}=e^{-im\int_{\mathcal{C}}d\tau  \sqrt{\dot{x}^\mu(\tau) \dot{x}^\nu(\tau) g_{\mu\nu}(x(\tau))}},
\end{equation}
for which the holonomy is proper length. According to the prescription in \eqref{LocalCorrelatorPrescription}, correlators of this holonomy can be extracted by taking derivatives with respect to mass.

\section{Length Correlators From Local Correlators}
\label{section3}

Correlators of massive quantum fields encode the on-shell action of massive point particles. For example, the on-shell action $\bar{S}$ of a point particle of mass $m$ is given by $\bar{S} = -\log \braket{\phi(x_1)\phi(x_2)}$ in the semiclassical regime, where the scalar field $\phi$ has mass $m$. This can be understood as a generalization of the geodesic approximation to non-inertial worldlines, and is made manifest in perturbation theory by the worldline representation of two-point functions. In this section, we use this idea to extend the prescription in Section \ref{section2} to correlators of local fields.

We propose that the prescription for obtaining $\braket{L_1 \cdots L_n}$ from $\bar{S}$ can be lifted to correlation functions of massive scalars $\phi_i$ in the semiclassical regime. Along the way, we discuss to what extent this local correlator prescription furnishes a fully quantum definition of length correlators. We then apply the correlator prescription to boundary correlators in AdS/CFT. We make use of the large body of sophisticated techniques for studying Witten diagrams to obtain closed-form expressions for proper length and correlators of other worldline observables in the bulk. 

\subsection{Local Correlators as Generating Functions of Length Correlators}
\label{CorrelatorLift}

The worldline representation of correlation functions recasts correlators of local operators as worldline path integrals, which provides intuition for how to extend the prescription in Section \ref{section2} to correlators of local operators. The vacuum two-point function of a scalar field $\phi$ of mass $m$ coupled to a second scalar $\psi$ can be written as
\begin{equation}
\braket{0|\phi(x_1)\phi(x_2)|0} = \int_0^\infty dT \int \mathcal{D}[x,\psi] e^{-\frac{1}{\hbar} S[x,\psi]} \equiv Z(x_1,x_2),
\end{equation}
where $S$ is the action of a particle of mass $m$ coupled to a scalar $\psi$. More explicitly, fixing $\psi$ to be a non-dynamical background field for simplicity, the path integral for $x$ is defined as follows,\footnote{This is analogous to the difference between Nambu-Goto and Polyakov form of the string action. In string theory, it is similarly common to define the action in Polyakov form.}
\begin{equation}
 Z(x_1,x_2) = \int_0^\infty dT \int_{x(0) = x_1}^{x(T) = x_2} \mathcal{D}[x]e^{-\frac{1}{\hbar}\frac{1}{2}\int_0^T d\tau \left( \dot{x}^\mu \dot{x}^\nu \eta_{\mu\nu} + m^2 + \lambda T^{-1} f(\psi) \right) }.
\end{equation}
One can obtain this form via Schwinger parametrization, in which case $T$ is the Schwinger parameter. Integrating over $T$ enforces reparametrization invariance. In the semiclassical regime, the saddle-point approximation to the $T$ integral recovers the square-root action \eqref{WorldlineActionSingle},
\begin{equation}
\hbar\rightarrow 0: ~~~~  Z(x_1,x_2) \rightarrow \int_{x(0) = x_1}^{x(1) = x_2} \mathcal{D}[x]e^{-\frac{1}{\hbar}\int_0^1 d\tau \left( m \sqrt{\dot{x}^\mu \dot{x}^\nu \eta_{\mu\nu}} +\lambda f(\psi) \right) }.
\label{2ptWorldlineNG}
\end{equation}
When $\psi$ is a dynamical quantum field, $S_{\psi}$ and the path integral over $\psi$ are included in the action above. Subleading corrections to the saddle-point approximation contain additional factors of $\hbar$. Therefore, to leading order in $\hbar$,\footnote{As elsewhere, we will sometimes omit overall factors of $\hbar$ but they can be easily restored and do not change our conclusions. We mention these factors only when relevant.}
\begin{equation}
\hbar \rightarrow 0:~~~~~~- \frac{d}{dm} \log Z(x_1,x_2)  \rightarrow \frac{d}{dm} \bar{S} = \braket{L(x_1,x_2)},
\label{sec3twoptproposal}
\end{equation}
which extends the prescription in Section \ref{section2} for the on-shell action to the two-point function in the semiclassical regime. From now on, we will work in the semiclassical regime unless specified otherwise. In a free theory, \eqref{sec3twoptproposal} recovers the standard geodesic approximation to the two-point function in the large-mass limit. 

We must include a dimensionful factor $\delta$ to render the argument of the logarithm dimensionless. If $\delta$ is a length scale, the appropriate quantity is $\log \left(Z(x_1,x_2)\delta^{2\Delta_0}\right)$ with $\Delta_0=(D-2)/2$. Natural choices for $\delta$ are $1/m$ and $\sqrt{x_{12}^2}$. Note that both contribute terms to $\braket{L}$ that are subleading in the semiclassical limit compared to $\bar{S}$. We therefore leave $\delta$ implicit until we discuss AdS/CFT, in which $\delta$ has a physical interpretation.

We propose that we similarly extract $\braket{L_1 \cdots L_n}_c$ for worldlines $1 \ldots n$ from higher-point correlators of $n$ fields $\phi_i$,
\begin{equation}
\braket{L_1(x_1,x_2) \cdots L_n(x_{2n-1},x_{2n})}_c = \prod_{i=1}^n \left(- \frac{d}{dm_i} \right) \log Z(x_1,\cdots,x_{2n}),
\label{LocalCorrelatorPrescription}
\end{equation}
with
\begin{align*}
Z(x_1,\cdots,x_{2n})
\equiv &
\braket{\phi_1(x_1)\phi_1(x_2)\cdots\phi_n(x_{2n-1})\phi_n(x_{2n})}
\\
=&
\prod_{i=1}^n \int_0^\infty dT_i \int \mathcal{D}[x_i,\psi] e^{-\frac{1}{\hbar} S[x_i,\psi]}.
\numberthis
\end{align*}
The worldline representation, written schematically above, can be obtained in perturbation theory by beginning with the field theory Feynman diagrams and then writing each free propagator in worldline representation. See \cite{Schubert01,StrasslerThesis,Strassler92} for review and explicit expressions for worldline networks, although we will not use these here. Unless specified otherwise, we will consider theories in which $\phi_i$ interact only with the field $\psi$ via vertices of the form $\lambda_i \phi_i^2 f_i(\psi)$ but have no self-interaction. These are the same theories whose worldline representations we have been studying \cite{StrasslerThesis}.

Although we work in the semiclassical regime, the definition of $\braket{L_1 \cdots L_n}_c$ in terms of a path integral further clarifies the difference between connected and disconnected correlators of $L_i$. Consider
\begin{equation}
Z=\braket{\phi_1(x_1)\phi_1(x_2) \phi_2(x_3) \phi_2(x_4)}.
\end{equation}
We have already identified
\begin{equation}
-\frac{d}{dm_1}\log Z = \braket{L_1(x_1,x_2)}
\end{equation}
as the one-point function of the length of worldline 1 in the presence of worldline 2, and similarly for $-\frac{d}{dm_2}\log Z$. It is natural to use the standard definition of the full correlator as
\begin{equation}
\braket{L_1(x_1,x_2)L_2(x_3,x_4)} \equiv Z^{-1}\frac{d}{dm_1}\frac{d}{dm_2}Z,
\end{equation}
which gives
\begin{align*}
\braket{L(x_1,x_2)L(x_3,x_4)}_c
=&
\frac{d}{dm_1}\frac{d}{dm_2} \log Z(m_1,m_2)
\\
=&
\braket{L(x_1,x_2)L(x_3,x_4)}
-
\braket{L(x_1,x_2)}\braket{L(x_3,x_4)},
\numberthis
\end{align*}
as expected. At higher points,
\begin{equation}
\braket{L_1(x_1,x_2) \cdots L_n(x_{2n-1},x_{2n})} \equiv Z^{-1}(x_1,\cdots,x_{2n}) \prod_{i=1}^n \left(- \frac{d}{dm_i} \right) Z(x_1,\cdots,x_{2n}).
\end{equation}

As an aside, another object associated with higher-point processes is the length of a worldline network. If we allow for self interactions of $\phi$, for instance a $\phi^3$ vertex, the length of a worldline network ending at multiple points $x_i$ appears as a tree-level contribution to
\begin{equation}
\braket{L(x_1,\cdots,x_4)} = -\frac{d}{dm} \log \braket{\phi(x_1) \phi(x_2) \phi(x_3) \phi(x_4)}.
\end{equation}
A similar prescription applies to the on-shell action.

Finally, while not the focus of this work, we note that correlators of $\phi_i$ appear to give a path-integral definition of correlators of length and other worldline observables that is valid in the fully quantum regime. The worldline representation tells us that local correlators can in practice be written as a sum over all possible worldlines connecting the local operator locations. Worldline observables in the classical theory capture what an observer measures along their on-shell worldline, but in this quantum case, they describe what a necessarily delocalized quantum observer measures. This basic idea is not very different from defining correlators or scattering processes on a background and including a path integral over the background configurations. In our case, the background is the worldline, and the boundary conditions are dictated by the local operator insertions. Investigating whether these boundary conditions can be interpreted as projective measurements of the initial and final states, or arise from Wightman, time-ordered, or in-in correlators may be relevant for connecting this proposal with Lorentzian setups, including the thought experiment described in the introduction.

As a concrete example, the expectation value of the interaction term $f(\psi(x))$ is
\begin{equation}
- \frac{d}{d\lambda} \log Z(x_1,x_2) = \frac{1}{Z}\int_0^\infty dT \int \mathcal{D}[x,\psi] \left( \int d\tau f(\psi(x(\tau))) \right) e^{-\frac{1}{\hbar} S[x,\psi]} \equiv \braket{f(\psi,x_1,x_2)}_{P},
\end{equation}
where in the semiclassical regime
\begin{equation}
\hbar \rightarrow 0: ~~~~~ \braket{f(\psi,x_1,x_2)}_{P} \rightarrow \int_{\gamma} d\tau f(\bar{\psi}(\bar{x}(\tau))),
\end{equation}
the integral of the on-shell value of $\bar{\psi}$ along the worldline $\gamma$ that connects $x_1,x_2$. The subscript $P$ on $\braket{f(\psi,x_1,x_2)}_{P}$ denotes averaging over worldline paths connecting $x_1,x_2$ as defined by the worldline path integral. The additional arguments $(x_1,x_2)$ denote the fact that this quantity is a function of $x_1,x_2$ and not any particular worldline or intermediate position.

In the fully quantum case, the QFT path integral representation of the two-point function implies that
\begin{equation}
\hbar \neq  0: ~~~~~\braket{f(\psi,x_1,x_2)}_{P} = \frac{\braket{\phi(x_1) \phi(x_2) \int d^Dx \phi^2(x) f(\psi(x))}}{\braket{\phi(x_1)\phi(x_2)}}.
\end{equation}
Similar statements apply to higher-point correlators.

A subtlety may occur when acting with $\frac{d}{dm}$ instead of $\frac{d}{d\lambda}$, because the path integral measure depends on $m$.\footnote{We thank J. Wilson-Gerow for discussions on this.} It turns out that in examples we check, the additional term is comparatively suppressed in the $\hbar \rightarrow 0$ limit. Modulo this subtlety, $d/dm$ inserts what may be interpreted as $\int ds$ in the worldline path integral, or $m\int d^Dx \phi^2$ in the $\phi$ path integral. In this sense, we see that $Z(x_1,x_2)$ may give a path-integral definition of length correlators in a fully quantum regime.

Pursuing the relation between first and second-quantized descriptions of these correlation functions may clarify to what extent these quantities can be interpreted as operators acting on a Hilbert space. Notably, these quantities are all integrated operators, and so for timelike worldlines in Lorentzian signature, are non-local in time. Finally, we note that using the einbein representation of the path integral in this context may be better for identifying proper length as an analytic function of worldline operators $\hat{x}$.

\subsection{QFT Examples}
\label{CorrelatorExamples}

In this section, we give some elementary examples of the local correlator proposal in \eqref{LocalCorrelatorPrescription}. The free scalar propagator in flat space is
\begin{equation}
G(x_1,x_2) = C_D ( m/ \sigma(x_1,x_2) )^{\Delta_0} K_{\Delta_0}( m \sigma(x_1,x_2)),
\end{equation}
where $C_D$ is an $m$-independent constant that depends on $D$ and $\sigma(x_1,x_2)$ is the geodesic distance $(\sigma(x_1,x_2))^2 = x_{12}^2$. $K_{\Delta_0}$ is a modified Bessel function of the second kind. The argument of $K_{\Delta_0}$ is $m \sigma(x_1,x_2)/\hbar$, and so the classical limit is the $m \sigma(x_1,x_2) \rightarrow \infty$ limit. $G(x_1,x_2)$ contains only two scales and therefore only one dimensionless ratio. Applying \eqref{LocalCorrelatorPrescription},
\begin{equation}
\braket{L(x_1,x_2)} = -\lim_{m \rightarrow \infty}\frac{d}{dm}\log G(x_1,x_2) =  \sigma(x_1,x_2).
\end{equation}
This is the familiar geodesic approximation, usually stated as $G(x_1,x_2)\sim e^{-m \sigma(x_1,x_2)}$ for $m \sigma(x_1,x_2) \rightarrow \infty$. The quantity above is slightly more refined in that it is an exact equality and $\sigma(x_1,x_2)$ is recovered with prefactor $1$.

We can also check a similar statement in AdS$_{d+1}$ with $D=d+1$. We use Poincare coordinates,
\begin{equation}
ds^2 = \frac{dz^2 + \delta_{ij} dx^i dx^j}{z^2},
\end{equation}
with $y^\mu = (z,x^i)$ and work in units of the AdS radius, $l_{AdS} = 1$. The bulk-to-bulk propagator is
\begin{equation}
G_\Delta(y_1,y_2) = \mathcal{C}_\Delta e^{-\Delta \sigma(y_1,y_2)}\F(\Delta, d/2,\Delta+1-d/2,e^{-2\sigma(y_1,y_2)}),~~~~~~~
\mathcal{C}_\Delta = \frac{2 \pi^{d/2}\Gamma(\Delta)}{\Gamma(\Delta-\Delta_0)},
\end{equation}
where mass and scaling dimension $\Delta$ are related by $m^2=\Delta(\Delta-d)$. In the large-$\Delta$ limit, $m \approx \Delta$, $\mathcal{C}_\Delta \approx \Delta^{\Delta_0}$ and for $d$ even, $\F(\Delta, d/2,\Delta+1-d/2,z) \approx g(z) +\mathcal{O}(1/\Delta)$ for some function $g$. Therefore in the geodesic approximation $\Delta \gg 1$,
\begin{equation}
\lim_{\Delta \rightarrow \infty} \braket{L(x_1,x_2)} = -\lim_{\Delta \rightarrow \infty} \frac{d}{d\Delta} \log G_\Delta(y_1,y_2)= \sigma(y_1,y_2).
\end{equation}
The same result is obtained from the propagator in arbitrary dimensions, whose large mass limit can be found in \cite{Maxfield17}.

Next, we consider higher-point processes. For free fields $\phi_i$, 
\begin{equation}
\braket{L_1(x_1,x_2)L_2(x_3,x_4)}_c =\frac{d^2}{dm_1 dm_2}\log \braket{\phi_1(x_1) \phi_1(x_2)\phi_2(x_3) \phi_2(x_4)} = 0,
\end{equation}
and when $x_i$ are distinct,
\begin{align*}
\lim_{m\rightarrow \infty} &\braket{L(x_1,\cdots,x_4)} = -\lim_{m\rightarrow \infty} \frac{d}{dm} \log \braket{\phi(x_1) \phi(x_2) \phi(x_3) \phi(x_4)} 
\\
=&\text{Min}(\sigma(x_1,x_2)+\sigma(x_3,x_4),\sigma(x_1,x_3)+\sigma(x_2,x_4),\sigma(x_1,x_4)+\sigma(x_2,x_3)).
\numberthis
\end{align*}

Next we consider a higher-point process in an interacting theory. Consider a three-point tree diagram in $\lambda \phi^2 \psi$ theory,
\begin{equation}
\braket{\phi(x_1)\phi(x_2)\psi(x_3)}= \lambda \int d^D x G_\phi(x_1,x)G_\phi(x_2,x)G_\psi(x_3,x).
\end{equation}
If every propagator were finite everywhere in the region of integration, we could take $m_\phi$ large and apply the geodesic approximation to $G_\phi(x_i,x)$. We would find that
\begin{equation}
\braket{\phi(x_1)\phi(x_2)\psi(x_3)} \sim \lambda \int d^D x e^{-m_\phi(\sigma(x_1,x)+\sigma(x_2,x))}G_\psi(x_3,x),
\end{equation}
up to polynomial functions in the integrand. The $x$ integral localizes onto its extremal value, which is a geodesic connecting $x_1,x_2$. With $m_\psi$ also large, the saddle then computes the minimum length of a geodesic network, which is the naive expectation from the proposal \eqref{LocalCorrelatorPrescription} in the semiclassical regime.

However, it is not immediately obvious the geodesic approximation applies for any large but finite values of $m_\phi, m_\psi$, because the coincident-point singularities of $G_\phi, G_\psi$ lie within the region of integration. We can however regulate these singularities by including a position-space cutoff, by for example excising balls around $x_i$ of some radius, or imposing a large-momentum cutoff. We can then apply the geodesic approximation safely, and it is then clear how \eqref{LocalCorrelatorPrescription} mechanically recovers the expected on-shell worldline action $\bar{S}$ at tree level in the large mass limit.

Rather than studying this further in flat space, we will carry out essentially this procedure for Witten diagrams in AdS/CFT. Witten diagrams have been studied extensively and are known in many cases to be finite for generic operator locations. In AdS/CFT, we will find that the proposal \eqref{LocalCorrelatorPrescription} does give the expected results in some cases.

\subsection{Application: AdS Worldline Observables From CFT Correlators}
\label{AdSCFT}

In the previous sections, we discussed how correlators of local operators encode the on-shell action of massive particles. Applying this idea to AdS/CFT, we might expect that CFT correlators encode an AdS observer's experience along its worldline in some simple way. An advantage of this approach to bulk reconstruction is that it would recover worldline observables in the familiar language of effective field theory as applied to correlators. This approach may also grant the boundary direct access to correlators of these worldline observables. The geodesic approximation is the standard method of extracting bulk proper time from boundary correlators, but generalizing to the more complex case of non-inertial bulk observers may help us identify a robust CFT dual of bulk proper time. Computing worldline observables in QFT is mechanical and involves computing worldline Feynman diagrams, so we may wish to locate the dual of this procedure in the CFT; following this same philosophy for Witten diagrams has already been particularly successful, after all. The foundation of this approach was laid in \cite{Maxfield17} by developing the AdS$_{d+1}$ worldline formulation of CFT$_d$ correlators.

By contrast, other approaches to bulk reconstruction appear less well-suited to computing worldline observables in interacting theories. Operators at a specific bulk point may require adding dressing (for example, see \cite{LewkowyczTV16}), while operators integrated along boundary-anchored worldlines may be easier to render diffeomorphism-invariant. It is currently unclear how to easily perform effective field theory computations in the language of quantum information or algebraic approaches to bulk reconstruction. The Hamilton-Kabat-Lifschytz-Lowe (HKLL) method of bulk reconstruction \cite{HKLL} has mostly been studied when the bulk is a free theory, and while the interacting version can be implemented in principle, computations are challenging and scarce (although see \cite{AnandCFKL17}). 

In this section, we apply a version of our proposal \eqref{LocalCorrelatorPrescription} to CFT correlators in order to extract bulk worldline observables. We focus on proper length but also obtain correlators of other integrated worldline quantities. We consider correlators of CFT single-trace primary operators $\Oc_i$ dual to AdS fields $\phi_i$. 

We find that bulk worldline observables are extracted with relatively little effort from Witten diagrams. This method provides a new use for the wealth of technology developed to study these diagrams at high loop order: one can simply take the large-$\Delta$ limit of existing results and then possibly take derivatives. Ultimately, this ease of use suggests that this approach may indeed be an efficient way to study observer-centric forms of bulk reconstruction.

\subsubsection{Two-Point Function}

We first extract the length of boundary-anchored geodesics from the CFT two-point function. 
\begin{equation}
\braket{L^{CFT}(x_1,x_2)} \equiv -\lim_{\Delta \rightarrow \infty }\frac{1}{2}\frac{d}{d\Delta} \log \braket{\Psi|\Oc_\delta (x_1) \Oc_\delta(x_2)|\Psi}= -\lim_{\Delta \rightarrow \infty }\braket{L(x_1,x_2)}_{AdS},
\end{equation}
where $\Oc_\delta(x_1)$ is a UV-regulated operator. The regulator $\delta$ corresponds to the location of the bulk cutoff surface that is necessary to render the worldline length finite. We will treat this regulator somewhat crudely, and simply take $\Oc_\delta(x_1) = \Oc(x_1)(\delta(x_1))^{\Delta}$ for simplicity. A more rigorous treatment may involve using HKLL to move the operator into the bulk. Taking $\ket{\Psi} = \ket{0}$,
\begin{align*}
\braket{L^{CFT}(x_1,x_2)}&=
-\frac{1}{2}\frac{d}{d\Delta} \log \left( \braket{\Oc (x_1) \Oc(x_2)}(\delta(x_1) \delta(x_2))^\Delta\right)
\\
&
=\log\left(\frac{|x_{12}|}{\sqrt{\delta(x_1)\delta(x_2)}}\right)=\braket{L(x_1,x_2)}_{AdS},
\numberthis
\end{align*}
the regulated length of a boundary-anchored geodesic. The precise relationship between bulk IR cutoff $\delta_i$ and boundary UV cutoff is somewhat ambiguous, but note that the $\delta \rightarrow 0$ limit is the geodesic limit for bulk propagators. We will leave the cutoff implicit going forwards.

Recovering the Euclidean bulk geodesic distance here was trivial. However, we briefly note an interesting feature. Taking $x_1=(i t_1,\vec{x}_1)$, $x_2=(i t_2,\vec{x}_2)$, and continuing $x_1,x_2$ to timelike separations, we cross the branch cut in the logarithm and find
\begin{equation}
\braket{L^{CFT}(x_1,x_2)} = \log |x_{12}| + \frac{i\pi}{2}.
\end{equation}
A priori, it is not clear if this corresponds to the length of some geodesic connecting the two boundary points, as there are no everywhere-timelike geodesics connecting timelike-separated points on the boundary. However, remarkably, $\braket{L^{CFT}(x_1,x_2)}$ agrees with the length of the novel mixed spacelike-timelike geodesic identified in AdS$_3$ in \cite{DoiHMTT22,DoiHMTT23}. The $\log |x_{12}|$ term corresponds to the length of boundary-anchored spacelike geodesics that extend into the bulk and the $i \pi/2$ is the length of a timelike geodesic that connects the two spacelike geodesics. 

In general dimensions, \cite{DoiHMTT22,DoiHMTT23} showed that boundary pseudo-entropy is computed by these mixed timelike-spacelike codimension-two surfaces. In our case, however, the length is associated with a one-dimensional curve, providing a different reconstruction of bulk proper time at least in this simple example. It would be interesting to explore these piecewise-geodesic curves in AdS$_{d+1}$/CFT$_{d}$ more generally, and also determine if they arise in flat space. The physical interpretation of these curves is not immediately clear, in particular whether the spacelike segments can be interpreted as tunnelling to a classically inaccessible region.

\subsubsection{Four-Point Function}

We now extract correlators of worldline operators from a boundary four-point function. We obtain these quantities from the exchange Witten diagram contribution to this four-point function. Unless otherwise specified, we will use the conventions in \cite{MeltzerPS19}, to which we refer the reader for further details and explicit expressions. 

We will use the pairwise-identical correlator $\braket{\Oc_1(x_1)\Oc_1(x_2)\Oc_2(x_3)\Oc_2(x_4)}$ of single-trace primaries $\Oc_1,\Oc_2$ dual to bulk scalars $\phi_1, \phi_2$. We consider a bulk theory with couplings $\lambda_1 \phi_1^2 \psi, \lambda_2 \phi_2^2 \psi$, for which only the $s$-channel exchange diagram contributes to $\braket{\Oc_1(x_1)\Oc_1(x_2)\Oc_2(x_3)\Oc_2(x_4)}$ at tree level in the bulk. The correlator can be written as a sum indexed by the contributions of primary operators $\Oc$,
\begin{equation}
\braket{\Oc_1(x_1)\Oc_1(x_2)\Oc_2(x_3)\Oc_2(x_4)} = T_s(x_i)\sum_{\Oc}
C_{\Oc_1\Oc_1\Oc}C_{\Oc_2\Oc_2\Oc} g_{\Delta,J}^{1122}(z,\bar{z}),
\end{equation}
with kinematic prefactor $T_s(x_i)$, OPE coefficients $C_{\Oc_i\Oc_j\Oc_k}$, and conformal blocks $g^{1122}_{\Delta,J}(x_i)$, which are functions of conformal cross-ratios $z,\bar{z}$ and independent of $\Delta_1, \Delta_2$. 

In the limit of heavy external dimensions, we will assume that the total contribution of double trace exchanges and certain derivatives thereof are suppressed with respect to that of the single-trace operator in some regime. We will be agnostic about the full regime of validity, other than arguing the assumption is valid for certain kinematics.

This assumption is justified as follows. In the conformal block decomposition, the exchanged operators are double traces $\Oc = [\Oc_1 \Oc_1]_{n,0},[\Oc_2 \Oc_2]_{n,0}$ and the single-trace operator $\Oc = \Oc_\psi$ dual to bulk $\psi$. We have checked that the product of OPE coefficients for the double trace operators, $C_{\Oc_1\Oc_1[\Oc_1\Oc_1]_{n,0}}C_{\Oc_2\Oc_2[\Oc_1\Oc_1]_{n,0}}$ for example, decays as a negative power of $\Delta_1$, $n$ at large $\Delta_1$ and large $n$ respectively. The derivative $d/d \Delta_1$ of this quantity also decays at large $\Delta_1,n$ as negative powers thereof. Conformal blocks decay exponentially at large exchange dimension $\Delta_e$ according to $z^{\Delta_e}$.\footnote{This suppression is what allows us to close the contour in the principal series integral representation of the four-point function and recover the conformal block expansion.} Putting these features together, we see that for $z, \bar{z} \ll 1/2$, the single-trace contribution dominates that of any other operator in the $\Delta_1,\Delta_2 \gg \Delta_\psi$ limit, even after we differentiate via $\frac{d}{d\Delta_1}$. Due to the conformal block suppression, double-trace exchanges are exponentially suppressed compared to single-trace exchanges, and therefore this suppression also holds for the sum of all double-trace contributions. As a consistency check, note that the lightest operator appearing in the $t,u$ channels is $[\Oc_1 \Oc_2]_{n,0}$, whose contribution is finite in the $t,u$ OPE limits; therefore there is no enhancement expected in the $s$-channel due to an infinite sum over blocks. The suppression of the double traces here is also consistent with \cite{PappadopuloRER12, KrausS18,GeodesicWittenDiagrams}, which include estimates of the rate of OPE convergence.

Moving on, we approximate the correlator as the single-trace contribution,
\begin{equation}
\braket{\Oc_1(x_1)\Oc_1(x_2)\Oc_2(x_3)\Oc_2(x_4)} \approx 
C_{\Oc_1\Oc_1\Oc_p}C_{\Oc_2\Oc_2\Oc_p} T_s(x_i)g_{\Delta_\psi,0}^{1122}(x_i).
\end{equation}
Before extracting the values of worldline observables, we will show that the quantity above can be written as the two-point function of worldline observables. This observation is by no means new, but we present it here in a way that may make applications to worldline observables more obvious.

We begin by recalling that a conformal block is computed in the bulk by a geodesic Witten diagram \cite{GeodesicWittenDiagrams},\footnote{In \cite{GeodesicWittenDiagrams}, propagators are normalized without the $C_{\Delta}$ factor, so that $\braket{\Oc(x_1)\Oc(x_2)}$ has unit coefficient. We will not adopt this normalization.}
\begin{align*}
g_{\Delta_5,0}^{1234}(x_i) =& \frac{\prod_{i=1}^5 \mathcal{C}(\Delta_i,0)}{T_s(x_i)\beta_{512}\beta_{534}}\int_{\gamma_{12}, AdS} dy_{12} \int_{\gamma_{34}, AdS} d y_{34} 
\\
&K_{\Delta_1} (x_1,y_{12}) K_{\Delta_2}(x_2,y_{12}) G_{\Delta_5}(y_{12},y_{34}) 
K_{\Delta_3}(x_3,y_{34}) K_{\Delta_4}(x_4,y_{34}),
\numberthis
\end{align*}
where the integrals run over geodesics $\gamma_{12},\gamma_{34}$ that connect boundary points $x_1,x_2$ and $x_3,x_4$ respectively, and
\begin{equation}
\beta_{\Delta 34}=\frac{\Gamma\left(\frac{\Delta+\Delta_{34}}{2}\right)\Gamma\left(\frac{\Delta-\Delta_{34}}{2}\right)}{2\Gamma(\Delta)},
\end{equation}
with $\Delta_{ij} = \Delta_i-\Delta_j$.

We can further simplify the geodesic Witten diagram expression. Moving to embedding space\footnote{See \cite{Penedones16} for review of embedding space and notation.} and following the approach in \cite{DyerFS17, ChenCLKN19},
\begin{equation}
\int_{\gamma_{12}, AdS} dy_{12} K_{\Delta} (x_1,y_{12}) K_{\Delta}(x_2,y_{12}) f(y_{12},y_{34})
=\int_{-\infty}^\infty d\lambda
\frac{1}{(P_1 \cdot X(\lambda))^{\Delta}(P_2 \cdot X(\lambda))^{\Delta}} f(X(\lambda)),
\end{equation}
where 
\begin{equation}
X^A(\lambda) = \frac{e^{\lambda} P_1^A+e^{-\lambda}P_2^A}{P_{12}^{1/2}}
\label{EmbeddingSpaceGeodesic}
\end{equation}
specifies an AdS geodesic between boundary points $P_1, P_2$ parametrized by $\lambda$, and $f$ is an arbitrary function. Following the conventions of \cite{DyerFS17} in mostly plus signature, points $P_i$ live on the projective null cone in the $d+2$-dimensional Minkowski space and encode boundary points. $X$ is a bulk point. These points obey $P^2 = 0, X^2 = -1$. The bulk-to-boundary propagator is $K(x_1,x_2,z_2) = (-2 P\cdot X)^{-\Delta}$. Using $P_{12} = (P_1-P_2)^2$ and \eqref{EmbeddingSpaceGeodesic}, one can show that
\begin{equation}
\int_{\gamma_{12}, AdS} dy_{12} K_{\Delta} (x_1,y_{12}) K_{\Delta}(x_2,y_{12}) f(y_{12})
=
x_{12}^{-2\Delta}
\int_{-\infty}^\infty d\lambda
f(y_{12}(\lambda)).
\label{GeodesicIdentity}
\end{equation}
This identity is implicitly derived in \cite{DyerFS17,ChenCLKN19}, and here we have merely noted that it trivially generalizes to arbitrary $f(X)$. See also \cite{HijanoKPS15,Maxfield17} for detailed comparison between geodesic Witten diagrams and worldline networks.

As an aside, we emphasize that converting between three-point structures and geodesic structures is exceedingly simple using \eqref{GeodesicIdentity}. We therefore expect that \eqref{GeodesicIdentity} can be used to obtain a geodesic decomposition of loop diagrams, which may prove useful for studying correlators of geodesic operators at loop level. Specifically, applying the split representation to loop diagrams converts them into gluings of three-point structures. Applying \eqref{GeodesicIdentity} next turns the three-point structures into geodesic three-point structures. The procedure in \cite{MeltzerPS19} can be then used to represent the OPE decomposition as a sum over such diagrams. Obtaining geodesic Witten diagrams for loop diagrams may also prove useful beyond computing correlators of geodesic operators.

Continuing on, \eqref{GeodesicIdentity} implies
\begin{equation}
g_{\Delta_5,0}^{1122}(z,\bar{z}) = \frac{\prod_i \mathcal{C}(\Delta_i,0)}{T_s(x_i)\beta_{511}\beta_{522}} x_{12}^{-2\Delta_1}x_{34}^{-2\Delta_2}\int d\lambda_{12} d \lambda_{34} G_{\Delta_5}(y_{12}(\lambda_{12}),y_{34}(\lambda_{34})).
\end{equation}
The two-point function here is normalized as $\braket{\Oc_1(x_1)\Oc_1(x_2)}=(\mathcal{C}(\Delta_1,0))^{2}x_{12}^{-2\Delta_1}$, and so
\begin{align*}
\braket{\Oc_1(x_1)\Oc_2(x_2)\Oc_2(x_3)\Oc_2(x_4)} \approx &
C_{\Oc_1\Oc_1\Oc_\psi}C_{\Oc_2\Oc_2\Oc_\psi} 
\frac{\mathcal{C}(\Delta_\psi,0)}{\beta_{\Delta_\psi 11}\beta_{\Delta_\psi 22}} \braket{\Oc(x_1)\Oc(x_2)}\braket{\Oc(x_3)\Oc(x_4)}
\\
&~~~~~~
\times\int d\lambda_{12} d \lambda_{34} G_{\Delta_\psi}(y_{12}(\lambda_{12}),y_{34}(\lambda_{34})).\numberthis
\end{align*}
This shows that, in the $\Delta_1, \Delta_2 \gg \Delta_\psi$ limit, the correlator localizes onto a quantity proportional to the two-point function of geodesic operators $\psi(\gamma_{ij})$. This is consistent with the worldline description studied in Section \ref{section2}. This basic feature was previously understood as a consequence of equality between geodesic operators and OPE blocks \cite{CzechLMMS16a,CzechLMMS16b,ChenCLKN19}. 

Finally, we use the CFT version of the local correlator proposal to compute worldline observables and their correlators with $\Delta_1, \Delta_2 \gg \Delta_\psi$. The generating function is 
\begin{align*}
\log Z(x_i)=&\log \braket{\Oc_1(x_1)\Oc_1(x_2)\Oc_2(x_3)\Oc_2(x_4)} 
\\
\approx &\log \braket{\Oc_1(x_1)\Oc_1(x_2)}\braket{\Oc_2(x_3)\Oc_2(x_4)} 
+ \frac{\braket{\Oc_1(x_1)\Oc_1(x_2)\Oc_2(x_3)\Oc_2(x_4)}_c}{\braket{\Oc_1(x_1)\Oc_1(x_2)}\braket{\Oc_2(x_3)\Oc_2(x_4)}}.
\numberthis
\end{align*}
The OPE coefficients and explicit conformal blocks can be found for example in \cite{MeltzerPS19, HeemskerkPPS09, GeodesicWittenDiagrams}, and so we will simply quote final results. According to the local correlator proposal, the two-point function of geodesic operators is
\begin{align*}
\braket{\psi(\gamma_{12}) \psi(\gamma_{34})}_{AdS} &= \frac{d}{d\lambda_1}\frac{d}{d\lambda_2} \log Z(x_i) =  \frac{d}{d\lambda_1}\frac{d}{d\lambda_2}\frac{\braket{\Oc_1(x_1)\Oc_1(x_2)\Oc_2(x_3)\Oc_2(x_4)}_c}{\braket{\Oc_1(x_1)\Oc_1(x_2)}\braket{\Oc_2(x_3)\Oc_2(x_4)}}
\\
&=\partial_{\lambda_1}C_{\Oc_1 \Oc_1 \Oc_\psi}\partial_{\lambda_2}C_{\Oc_2 \Oc_2 \Oc_\psi}g^{1122}_{\Delta_\psi,0}(z,\bar{z}),
\numberthis
\end{align*}
where
\begin{equation}
C_{\Oc_1 \Oc_1 \Oc_\psi} C_{\Oc_2 \Oc_2 \Oc_\psi} 
\approx 
\lambda_1 \lambda_2 \frac{\pi ^d  \left(d-2 \Delta _\psi\right) \Gamma^4 \left(\frac{\Delta _\psi}{2}\right) \Gamma \left(\frac{d}{2}-\Delta _\psi\right)}{8 \Delta _1^{\frac{d}{2}} \Delta _2^{\frac{d}{2}} \Gamma^2 \left(\Delta _\psi\right) \Gamma \left(\frac{d}{2}-\Delta _\psi+1\right)}.
\end{equation}

We can also take the derivative with respect to $\Delta_1$ to find our predicted correction to the proper length of a particle following geodesic $\gamma_{12}$ due to a $\psi$-mediated force from the particle following $\gamma_{34}$,
\begin{equation}
\braket{L(x_1,x_2)}_{AdS}= \log |x_{12}|+
\lambda_1 \lambda_2 \frac{d \pi ^d \left(d-2 \Delta _\psi\right) \Gamma^4 \left(\frac{\Delta _\psi}{2}\right) \Gamma \left(\frac{d}{2}-\Delta _\psi\right)}{32 \Delta _1^{\frac{d}{2}+1} \Delta _2^{\frac{d}{2}} \Gamma^2 \left(\Delta _\psi\right) \Gamma \left(\frac{d}{2}-\Delta _\psi+1\right)}
g^{1122}_{\Delta_\psi,0}(z_i).
\end{equation}
We have argued that this correction to proper length corresponds to non-inertial motion in the bulk. The correction vanishes at large $\Delta_1$, which is a rudimentary consistency check. Another example one may explore is the tree-level contact diagram.

We see that it was straightforward to extract $\braket{L(x_1,x_2)}$ from the known OPE decomposition of Witten diagrams. We simply took the large $\Delta_i$ limit of existing expressions and then applied derivatives. As the conformal block was independent of external dimensions, only the OPE data were needed.

\subsubsection{AdS Length as CFT Correlation Function Entropy}
\label{CorrelatorEntropy}

Our CFT proposal for AdS length does not obviously arise from any known property of CFT. In fact, it is unclear if the CFT proposal is well-defined for generic CFTs. In writing $\frac{d}{d\Delta} \log \braket{\Oc(x_1)\Oc(x_2)}$, we have assumed that $\Oc$ belongs to a family of operators continuously parametrized by $\Delta$. This is reasonable for CFTs dual to QFTs in AdS in which the mass is a tunable parameter. However, generic CFTs may not contain a family of operators with this property.\footnote{In CFT, continuous quantum numbers can sometimes be made sense of formally \cite{LorentzianInversionFormula,KravchukS18}.} 

Nevertheless, in this section we will identify a quantity defined in quantum mechanics that, at least in certain simple cases, does define a family of $\Oc$ appearing in the CFT proposal. This section will be exploratory, and we will merely demonstrate a few basic features to motivate further investigation.

Our approach here superficially resembles methods used to study von Neumann entropy in AdS/CFT. Specifically, our present task is analogous to attempting to (re)discover that von Neumann entropy of subregions in the CFT is a boundary dual of extremal surfaces in AdS.

In quantum mechanics, we consider
\begin{equation}
S(\Oc,\rho) \equiv -\frac{\text{tr}  (\rho  \Oc \log \Oc)}{\text{tr}(\rho \Oc)} = - \frac{d}{dn}\bigg|_{n=1} \log \text{tr}  (\rho  \Oc^n),
\label{OperatorEntropy}
\end{equation}
whenever the quantities above are well defined. One example is when $\Oc$ is a Hermitian operator. The log operator is defined via the replica trick above.

For clarity, we refer to \eqref{OperatorEntropy} as operator entropy for a single $\Oc$ and correlation function entropy with multiple $\Oc$ insertions. This quantity is a generalization of von Neumann entropy in the specific sense that when $\rho =\mathbf{1}/\text{tr}(\mathbf{1})$ and $\Oc$ is a different density matrix, the resulting operator entropy is exactly the von Neumann entropy of $\Oc$. We also consider the related quantity
\begin{equation}
S^{geo}(\Oc,\rho) \equiv \left(1- \frac{d}{dn}\right) \log \text{tr}  (\rho  \Oc^n) \bigg|_{n=1},
\end{equation}
defined analogously to geometric entropy.

We can show that $S^{geo}(\Oc,\rho) = 0$ if $\rho$ is an eigenstate of $\Oc$, and generically is nonzero otherwise. In this sense, $S^{geo}(\Oc,\rho)$ partially distinguishes eigenstates and other states.\footnote{For comparison, von Neumann entropy distinguishes pure and mixed states.} For example, consider a two-level system in the normalized state $a_1 \ket{\psi_1}+a_2\ket{\psi_2}$ written in the eigenbasis of $\Oc$, where $\Oc = \text{diag}(\lambda_1,\lambda_2)$. Assuming $a_i$ are real for convenience,
\begin{equation}
S^{geo}(\Oc,a_1 \ket{\psi_1}+a_2\ket{\psi_2}) =\log  (a_1^2 \lambda_1 + a_2^2 \lambda_2 )-\frac{a_1 ^2\lambda_1\log \lambda_1 + a_2^2 \lambda_2 \log\lambda_2 
}{a_1^2 \lambda_1 + a_2^2 \lambda_2}
.
\end{equation}
However,
\begin{equation}
S^{geo}(\Oc,\ket{\psi_1}) =S^{geo}(\Oc,\ket{\psi_2})= 0.
\end{equation}
Essentially, $\braket{\psi_i|\Oc^n|\psi_i} = \lambda_i^n$ and $(1-d/dn)\log \lambda_i^n \big|_{n=1}  =0$.

Trivially, the same statement holds for eigenstates of products of operators. To illustrate, consider two systems whose joint Hilbert space is $\mathcal{H}_A \otimes \mathcal{H}_B$, where $\mathcal{H}_A, \mathcal{H}_B$ have dimensions $d_A, d_B$. In the eigenbases of operators $\Oc_A, \Oc_B$, which act nontrivially in $\mathcal{H}_A, \mathcal{H}_B$ respectively, consider the states $\ket{\psi_a} = (a_1,a_2, \cdots,a_{d_A}) \in \mathcal{H}_A$, $\ket{\psi_b} = (b_1,b_2,\cdots,b_{d_B}) \in \mathcal{H}_B$, and the product state $\ket{\psi_a \psi_b} \in \mathcal{H}_A \otimes \mathcal{H}_B$. Then,
\begin{align*}
S^{geo}&(\Oc_A\Oc_B, \ket{\psi_a \psi_b}) = \left(1-\frac{d}{dn} \right) \log \braket{\psi_a \psi_b|\Oc_1^{n} \Oc_2^{n} |\psi_a \psi_b}\bigg|_{n=1} 
\\
=&
\log \sum_i a_i^2 \lambda_{a,i} \sum_j b_j^2 \lambda_{b,j}-
\frac{
\sum_i a_i^2 \lambda_{a,i} \log \lambda_{a,i} \sum_j b_j^2 \lambda_{b,j}
+
\sum_i a_i^2 \lambda_{a,i} \sum_j b_j^2 \lambda_{b,j} \log \lambda_{b,j}
}{\sum_i a_i^2 \lambda_{a,i} \sum_j b_j^2 \lambda_{b,j}}.
\numberthis
\end{align*}
If we instead consider eigenstates, $\Oc_A \Oc_B \ket{\psi_a \psi_b} = \lambda_a \lambda_b \ket{\psi_a \psi_b}$, then 
\begin{equation}
S^{geo}(\Oc_A\Oc_B,\ket{\psi_a \psi_b}) = \ln (\lambda_a \lambda_b)-
\ln \lambda_a  - \ln \lambda_b = 0.
\end{equation}
More generally, it is clear that
\begin{equation}
S^{geo}(\Oc,\ket{\psi}) = 0 ~~\text{if}~~ \Oc \ket{\psi} = \lambda \ket{\psi} + \ket{\psi'}, ~~~ \braket{\psi |\psi'} = 0.
\label{OperatorEntropyZeroCondition}
\end{equation}
Above, $\ket{\psi}$ is an eigenstate up to additional states that when acted upon by $\Oc$ are orthogonal to $\ket{\psi}$. For clarity, we note that according to \eqref{OperatorEntropyZeroCondition}, non-vanishing geometric operator entropy implies the state was not an eigenstate,
\begin{equation}
S^{geo}(\Oc,\ket{\psi}) \neq 0 \implies \Oc \ket{\psi} \neq \lambda_\psi \ket{\psi}.
\end{equation}
We have not shown that $S(\Oc,\rho) = 0$ implies $\rho$ is an eigenstate, but it would be interesting to identify a quantity that obeys this stronger condition.

Correlator entropy can also be studied in QFT. In mean field theory, consider the normal-ordered $\Oc^n(x)$ normalized as $\braket{0|\Oc^n(x_1)\Oc^n(x_2)|0} $ $=$ $ \braket{0|\Oc(x_1)\Oc(x_2)|0}^n$.\footnote{One may also consider $n$-insertions at slightly different points so that $\mathcal{O}^n$ does not require normal ordering.} Then,
\begin{equation}
S(\Oc(x_1)\Oc(x_2),\ket{0})= \Delta \log(|x_{12}|)= \Delta \braket{L(x_1,x_2)}_{AdS}=\Delta \braket{L^{CFT}(x_1,x_2)}.
\end{equation}
Similar relations hold for higher-point functions in mean field theory, including statements similar to those in Section \ref{CorrelatorExamples}. 

For comparison, $S^{geo}(\Oc(x_1)\Oc(x_2),\ket{0})=0$. For a free massive scalar $\phi$, we have $S^{geo}(\phi(x_1)\phi(x_2),\ket{0}) =0$ as well, which we can understand as follows. Expanding $\phi(x_1) \phi(x_2)$ in $a_k, a^\dagger_k$ operators and commuting $a_k$ to the right, we see how this example satisfies the condition \eqref{OperatorEntropyZeroCondition}. $\phi(x_1)\phi(x_2)$ contains the identity operator, of which the vacuum is an eigenstate, and other operators that have no overlap with the vacuum, for example operators that create two particles $a^\dagger_{k_1} a^\dagger_{k_2}$.

Finally, we note $\Delta \frac{d}{d\Delta}$ and $\frac{d}{dn}\big|_{n=1}$ are equivalent in a certain sense in the context of Witten diagrams. We can see this via an observation reminiscent of \cite{LewkowyczM13}. In AdS, $\braket{\phi^n(y_1)\phi^n(y_2)} = G^n_\Delta(y_1,y_2) \neq G_{n\Delta}(y_1,y_2)$ and so for bulk fields, the actions of $\Delta \frac{d}{d\Delta}$ and $\frac{d}{dn}\big|_{n=1}$ are entirely different. However, when we take one of the points to the boundary, $y_2=(x_2,\delta_2)$, 
\begin{equation}
G_\Delta^n(y_1,x_2,\delta_2) \approx \delta_2^{n\Delta} K^n_\Delta(y_1,x_2)=\delta_2^{n\Delta} K_{n\Delta}(y_1,x_2)\approx G_{n\Delta}(y_1,x_2,\delta_2),
\end{equation}
where we have used that $K_{\Delta}(y_1,x_2)$ is a power in $\Delta$, and kept only the leading term in the $\delta_2 \rightarrow 0$ limit.

While we have presented some quantum-mechanical curiosities in this section, it remains unclear whether there is a well-defined CFT dual to proper length.

\section{Future Directions}
\label{futures}

The main purpose of this work was to show that correlators of worldline proper length at tree level are computed by mass derivatives of the on-shell action. We found this prescription gives a straightforward algorithm to compute correlators of length, easily incorporates QFT effects, and furnishes Feynman diagrams for proper length correlators. We also provided evidence that the logarithm of local correlators serves as a generating function of worldline observable correlators. To our knowledge, $n$-point correlators of worldline proper length have not been studied in full generality. Nevertheless, the on-shell action is a well-studied object and we encountered no fundamental obstacles to studying length correlators in experiment, gravity, and holography. Length correlators therefore appear to be ripe for exploration. 

We conclude by highlighting specific future directions.

\subsection{On-Shell Action and QFT}

It would be interesting to compute proper length correlators directly from time operators suitably defined in the worldline quantum mechanics and compare to the prescription we presented. This may be a consistency check of our main result, and may also teach us more about proper time and length as operators in the fully quantum case. Motivated by our local correlation function proposal, one may also investigate length as an operator in second quantization. To this end, exploring coincident-point singularities of proper length correlators may be informative. We considered a model of an ideal clock, but studying a more realistic clock with finite energy resolution may have a number of applications to experiment and theory.

We chose worldlines of finite extent, but the on-shell action appears in descriptions of classical scattering processes. We expect that explicit computations of length correlators are easier in momentum space than in position space. It would be interesting to understand whether the S-matrix encodes proper length correlators. One may investigate the relation to the Detweiler redshift \cite{Detweiler08} and also time delays.

It would be interesting to explore proper length correlators in the fully quantum regime, for example by using correlators of local operators as the generating function, as we briefly discussed. One may investigate signatures of tunnelling processes in length correlators, and also whether positivity of length or the triangle inequality apply in some form. One may also include loop diagrams and investigate renormalization of length. 

We studied proper length for massive probes but whether this can be extended in some form to massless probes is unclear. For example, one may consider $\frac{d}{dm}\big|_{m=0}$, essentially as employed in \cite{LinMRS22a,LinMRS22b}. It would be interesting if observables in laser interferometers could be recast in terms of length correlators. In the standard approach, the interferometer observable in a certain gauge is derived from a two-point function of proper time of a massive worldline. When the worldline is coupled to a quantum field theory, this quantity may be related to the length correlators described here. However, here we fixed the endpoints at some coordinate value, which may not be sensible in gravity, and we worked primarily in Euclidean signature.

The methods presented here may be useful for computing correlators of certain worldline observables discussed in \cite{Witten23a,Witten23b} when the worldlines are non-inertial. We briefly studied similar correlators for geodesics in non-gravitational theories, but finite-$m$ corrections and perturbative quantum gravity effects can be included using standard worldline technology, as we showed. Proper time correlators may also be convenient objects to study here, if well defined. Explicit computations of simple worldline observables in QFT coupled to gravity may provide a useful testbed for recent ideas on algebras of observables, relationality, and dressing in gravity. Leading perturbative corrections to low-point correlators are an appealingly concrete target.

\subsection{Holography}

Despite recent attention, the emergence of bulk worldline proper time from the boundary warrants continued study, and in particular of how the boundary encodes the proper time of an infalling observer that reaches the black hole singularity. Following \cite{DoiHMTT22,DoiHMTT23} and our brief exploration here, mixed timelike-spacelike geodesics may also be interesting if they appear more generally. Deriving these geodesics from two-point functions may clarify their interpretation. It would also be interesting to extract bulk lengths in a similar way from the S-matrix, if possible. To this end, working in the language of a putative celestial dual or with Witten diagrams in an auxiliary AdS space may be helpful.

We expect that connecting our proposal with the length operator derived in \cite{Afkhami-JeddiHKT17} may be fruitful. This operator was derived in the Regge limit by using HKLL reconstruction to recast the integrated stress tensor on the boundary as the bulk operator $\int h$ integrated along a null trajectory. The procedure we employed works for more general kinematics, but is nevertheless in a certain sense an extension of the procedure in \cite{Afkhami-JeddiHKT17} to operators besides the stress tensor, as these also contribute to worldline length with generic kinematics. The approach in \cite{Afkhami-JeddiHKT17} may also be useful for refining our ad-hoc treatment of the cutoff, or computing correlators of bulk length operators in the appropriate Regge limits.

We identified rudiments of a possible CFT dual of proper length. Logarithmic operators were central to this construction, and so it may be useful to determine whether these log operators exist in generic CFTs, or to compare to properties of log operators in logarithmic CFTs (see \cite{HogervorstPV16} and references therein). Examples of CFTs with families of operators whose scaling dimensions depend on a continuous parameter may be useful for studying derivatives $\frac{d}{d\Delta}$ of correlation functions or operators. One may also investigate derivatives with respect to other quantum numbers such as spin.

We computed correlators of worldline observables via the large scaling dimension limit of Witten diagrams, which we expect recovers the geodesic approximation for bulk propagators. Additional computations would help explore this idea. Correlation functions of worldline observables, including proper length, provide a novel use for the large body of Witten diagram computations. It may also be more efficient to instead use worldline diagrams in AdS from the outset to compute the bulk on-shell action \cite{Maxfield17}. Bulk worldline computations may provide a non-trivial check of the CFT proposal for computing bulk worldline observables. Just as conformal symmetry has been a powerful tool for computing Witten diagrams, it would be interesting if it can be leveraged to further simplify AdS worldline computations.

Comparing the definition of length correlators presented here with recent proposals for length operators in lower-dimensional systems \cite{IliesiuLLMM24,AlmheiriGH24} may be fruitful. Two-dimensional gravity, SYK models, and AdS$_2$/CFT$_1$ holography differ in significant ways from the higher-dimensional setups we studied. Nevertheless, these ideas taken altogether may be useful for studying length correlators in higher dimensions. If our proposal can be adapted to these low-dimension settings, one may determine whether it satisfies all four criteria listed in \cite{IliesiuLLMM24} if applicable.

It would be interesting to further explore the connection between OPE blocks and bulk worldline operators as studied in \cite{CzechLMMS16a,CzechLMMS16b}. Continuing operators to timelike separations may also make contact with the mixed spacelike-timelike geodesics studied earlier \cite{DoiHMTT22,DoiHMTT23}, the entanglement first law generalized to OPE blocks \cite{deBoerHHM16}, and the length operator derived in \cite{Afkhami-JeddiHKT17} in the Regge limit. Notably, the stress tensor OPE block for operators at timelike separations is the vacuum-subtracted modular Hamiltonian \cite{CzechLMMS16a,CzechLMMS16b} and may encode a length operator for the associated mixed timelike-spacelike geodesics.

\section{Acknowledgements}

We thank Leonardo Badurina, Clifford Cheung, Elliott Gesteau, Flaminia Giacomini, Thomas Hartman, Temple He, Eliot Hijano, Philipp A. Hoehn, Joon-Hwi Kim, Per Kraus, David Meltzer, Sridip Pal, Julio Parra-Martinez, Ryan Plestid, David Simmons-Duffin, Jordan Wilson-Gerow, and Kathryn Zurek for discussions and/or comments on drafts. We also thank Temple He, Jordan Wilson-Gerow, and Kathryn Zurek for collaboration during an early stage of this work. AS is supported by the Heising-Simons Foundation “Observational Signatures of Quantum Gravity” collaboration grant 2021-2817, the U.S. Department of Energy, Office of Science, Office of High Energy Physics, under Award No. DE-SC0011632, and the Walter Burke Institute for Theoretical Physics.

\bibliographystyle{JHEP}
\bibliography{refs}
\end{document}